\documentclass[fleqn,usenatbib]{mnras}

\usepackage{newtxtext,newtxmath}

\usepackage[T1]{fontenc}

\DeclareRobustCommand{\VAN}[3]{#2}
\let\VANthebibliography\thebibliography
\def\thebibliography{\DeclareRobustCommand{\VAN}[3]{##3}\VANthebibliography}

\usepackage{graphicx}	
\usepackage{amsmath}	
\usepackage{longtable}
\usepackage{booktabs}

\title[\textit{TESS} Occurrence Rates by Spectral Type]{Demographics of Close-In \textit{TESS} Exoplanets Orbiting FGK Main-sequence Stars -- II. Spectral-Type Dependence and the Neptunian Landscape}

\author[K. Cui et al.]{
Kaiming Cui,$^{1,2}$\thanks{E-mail: Kaiming.Cui@warwick.ac.uk}
David J. Armstrong,$^{1,2}$
Andreas Hadjigeorghiou,$^{1,2}$
Marina Lafarga,$^{1,2}$
\newauthor
Ares Osborn,$^{1,2}$
George W. King,$^{1,2}$
Tim Hallatt,$^{3}$
Vedad Kunovac,$^{1,2}$
Lauren Doyle$^{1,2}$
\\
$^{1}$Department of Physics, University of Warwick, Gibbet Hill Road, Coventry CV4 7AL, UK\\
$^{2}$Centre for Exoplanets and Habitability, University of Warwick, Gibbet Hill Road, Coventry CV4 7AL, UK\\
$^3$MIT Kavli Institute for Astrophysics and Space Research, Massachusetts Institute of Technology, Cambridge, MA 02139, USA
}

\date{Accepted XXX. Received YYY; in original form ZZZ}

\pubyear{\the\year{}}

\begin{document}
\label{firstpage}
\pagerange{\pageref{firstpage}--\pageref{lastpage}}
\maketitle

\begin{abstract}
Understanding how planetary demographics depend on host star properties is essential for constraining planet formation and evolution theories. In this paper, we analyse close-in ($0.5$--$16$\,d) planet occurrence rates from the \textit{TESS} Science Processing Operations Center (SPOC) full-frame image data for individual F-, G-, and K-type main-sequence stellar samples.
Integrated over $0.5 < P < 16$\,d and $2 < R_\mathrm{p} < 20\,R_\oplus$, the occurrence rates are $12.1^{+3.0}_{-2.3}$, $10.2^{+1.2}_{-1.0}$, and $13.1^{+1.4}_{-1.2}$ planets per 100 stars for F-, G-, and K-type hosts, consistent with each other at $1\sigma$, so any spectral-type dependence is localized rather than global. For transiting hot Jupiters, we measure $0.30^{+0.04}_{-0.03}$, $0.50\pm0.03$, and $0.67^{+0.08}_{-0.07}$ per cent, with the rate peaking at K type; the latter being the first such measurement for K-type hosts. Modelling the Neptunian desert directly in occurrence-rate space, we find that the desert extends to longer orbital periods around hotter stars. Integrating over the desert interior gives $0.069^{+0.045}_{-0.011}$, $0.093^{+0.046}_{-0.022}$, and $0.150^{+0.033}_{-0.026}$ planets per 100 stars, implying about one close-in planet in a hundred lies inside the desert, and the desert is roughly twice as populated around K- as around F-type hosts. A rigorous Bayesian model comparison for the $6$--$8\,R_\oplus$ strip indicates that a distinct Neptunian ridge is not required to explain the measurements.
\end{abstract}

\begin{keywords}
methods: data analysis -- methods: statistical -- planets and satellites: detection -- planets and satellites: fundamental parameters -- stars: fundamental parameters
\end{keywords}



\section{Introduction}\label{sec:intro}

The demographic study of exoplanets has become one of the most powerful tools for constraining theories of planet formation and evolution \citep[e.g.][]{mordasiniExtrasolarPlanetPopulation2009, muldersExoplanetPopulationObservation2019, drazkowskaPlanetFormationTheory2023}. Transit and radial velocity surveys have revealed a rich structure in the orbital period--radius and period--mass planes, including prominent features such as the hot Jupiter pile-up \citep[e.g.][]{wrightFrequencyHotJupiters2012}, the Neptunian desert \citep[e.g.][]{szaboShortperiodCensorSubJupiter2011, beaugeEMERGINGTRENDSPERIOD2012, mazehDearthShortperiodNeptunian2016}, and the radius valley \citep[e.g.][]{fultonCaliforniaKeplerSurveyIII2017, vaneylenAsteroseismicViewRadius2018}. These structures are the imprints of physical processes such as core accretion \citep[e.g.][]{pollackFormationGiantPlanets1996, idaDeterministicModelPlanetary2004}, disc migration \citep[e.g.][]{kleyPlanetDiskInteractionOrbital2012, jinPLANETARYPOPULATIONSYNTHESIS2014}, atmospheric mass loss driven either by stellar irradiation \citep[e.g.][]{owenKeplerPlanetsTale2013,lopezROLECOREMASS2013,owenEvaporationValleyKepler2017} or by the cooling luminosity of the core \citep[e.g.][]{ginzburgCorepoweredMasslossRadius2018, guptaSculptingValleyRadius2019, guptaSignaturesCorepoweredMassloss2020}, and high-eccentricity migration followed by tidal evolution \citep[e.g.][]{rasioDynamicalInstabilitiesFormation1996, fabryckyShrinkingBinaryPlanetary2007, owenPhotoevaporationHigheccentricityMigration2018, dawsonOriginsHotJupiters2018}. 
Because these mechanisms can produce similar demographic features, distinguishing between them requires establishing how the features scale with host-star properties.

Theoretical models predict systematic variations in planet formation and evolution across stellar types, as disc masses and thermal structures, disc dispersal timescales, and irradiation-driven atmospheric evolution depend on host-star properties \citep[e.g.][]{kennedyPlanetFormationStars2008, ribasProtoplanetaryDiskLifetimes2015,pascucciSTEEPERLINEARDISK2016, owenEvaporationValleyKepler2017}. The dependence is often diagnostic: photoevaporation is set by the XUV luminosity whereas core-powered mass loss is set by the bolometric flux, so the two predict different scalings of the radius valley with stellar mass \citep[e.g.][]{guptaSignaturesCorepoweredMassloss2020, rogersPhotoevaporationCorepoweredMassloss2021}, and tidal dissipation weakens toward earlier spectral types, so close-in giants inspiral at different rates around F, G, and K hosts \citep[e.g.][]{hamerHotJupitersAre2019, barkerTidalDissipationEvolving2020}. Observationally, several studies using \textit{Kepler} data have found that the occurrence of small planets increases toward later spectral types \citep[e.g.][]{howardPLANETOCCURRENCE0252012, muldersINCREASEMASSPLANETARY2015, petiguraCaliforniaKeplerSurveyIV2018, kunimotoSearchingEntiretyKepler2020, hsuOccurrenceRatesPlanets2020}, and that the radius valley shifts to larger radii around more massive hosts \citep[e.g.][]{bergerGaiaKeplerStellarProperties2020a, petiguraCaliforniaKeplerSurveyRadius2022, hoDeepRadiusValley2023}. However, these studies are predominantly based on the narrow \textit{Kepler} field, and extending such analyses to the all-sky \textit{TESS} sample offers both a larger stellar baseline and access to brighter, better-characterized targets.

The spectral-type dependence of specific features in the period--radius plane is of particular interest. The Neptunian desert, the observed dearth of intermediate-sized planets at very short orbital periods, likely reflects multiple processes: atmospheric escape may shape its lower boundary, whereas high-eccentricity migration, tidal disruption, and tidal inspiral followed by Roche-lobe overflow may influence its upper boundary and populate the desert with stripped remnants \citep[e.g.,][]{mazehDearthShortperiodNeptunian2016, lundkvistHotSuperEarthsStripped2016, owenPhotoevaporationHigheccentricityMigration2018, vissapragadaUpperEdgeNeptune2022, hallattSheddingLightDesert2026}. The locations and prominence of these features may vary with spectral type because stellar irradiation histories \citep[e.g.][]{hallattSculptingSubSaturnOccurrence2022} and tidal dissipation \citep[][]{hallattRevealingOriginDesert2026} depend on host-star properties. 
{Likewise, hot-Jupiter occurrence appears to vary with stellar types \citep[e.g.,][]{zhouTwoNewHATNet2019, moeImpactBinaryStars2021, beleznayExploringDependenceHot2022, bryantOccurrenceRateGiant2023, ganOccurrenceRateHot2023}. 
Recent studies have also proposed a ridge near the desert edge and an adjacent savanna \citep[][]{bourrierDREAMOrbitalArchitecture2023, castro-gonzalezMappingExoNeptunianLandscape2024}.} Quantifying these structures separately across spectral types can therefore test competing formation and evolutionary pathways.
 
In Paper~I \citep[][]{cuiDemographicsCloseinTESS2026}, we presented the first comprehensive census of close-in ($0.5 < P < 16$\,d) planet occurrence rates around FGK main-sequence stars using four years of \textit{TESS}-SPOC full-frame image (FFI) data, combining a GPU-accelerated box least-squares (BLS) search of ${\sim}2.1$ million FGK dwarfs with a machine-learning vetting and validation chain. Because that census treated FGK stars as a single population, it could not separate genuine trends with host-star properties from the composition of the underlying stellar sample. 

{In this paper, we present spectral-type-resolved occurrence rates for close-in \textit{TESS} planets around F-, G-, and K-type main-sequence stars by dividing the sample into F, G, and K subsamples and recomputing the detection and vetting efficiencies for each, enabling consistent comparisons of occurrence rates across spectral type over the close-in period--radius plane.} In Section~\ref{sec:methods}, we describe our stellar subsamples by spectral type, the adaptations made to our completeness estimation, and our occurrence-rate models. In Section~\ref{sec:results}, we present weighted kernel density estimates (wKDEs) and two-dimensional occurrence-rate maps for each spectral type; model the upper and lower boundaries of the Neptunian desert; test the significance of the Neptunian ridge through one-dimensional occurrence-rate modeling; and measure the hot-Jupiter occurrence rate for each subsample. In Section~\ref{sec:discussion}, we compare our results with those from \textit{Kepler} and discuss the dependence of the desert boundaries on stellar type. Finally, we summarize our conclusions in Section~\ref{sec:conclusion}.

\section{Methods}\label{sec:methods}

The stellar sample, planet detection pipeline, vetting and validation procedures, and occurrence rate estimation framework used in this work are those established in Paper~I. Here we provide a brief summary and describe the adaptations made for spectral-type-resolved analyses.

\subsection{Stellar Sample}\label{subsec:sample}

Our parent sample consists of the FGK main-sequence stars from the \textit{TESS}-SPOC FFI catalogue covering Sectors 1--55, as described in Paper~I and \citet{doyleTESSSPOCFFITarget2024}. We require {$G < 14$\,mag}, \texttt{parallax\_over\_error} $> 5$, and $\log g > 3.5$, and restrict effective temperatures to $3900$--$7300$\,K based on {TIC~v8.2} \citep{stassunRevisedTESSInput2019, paegertTESSInputCatalog2021}. This yields 2,097,359 FGK targets as a magnitude-limited sample.

\begin{figure}
    \centering
    \includegraphics[width=\columnwidth]{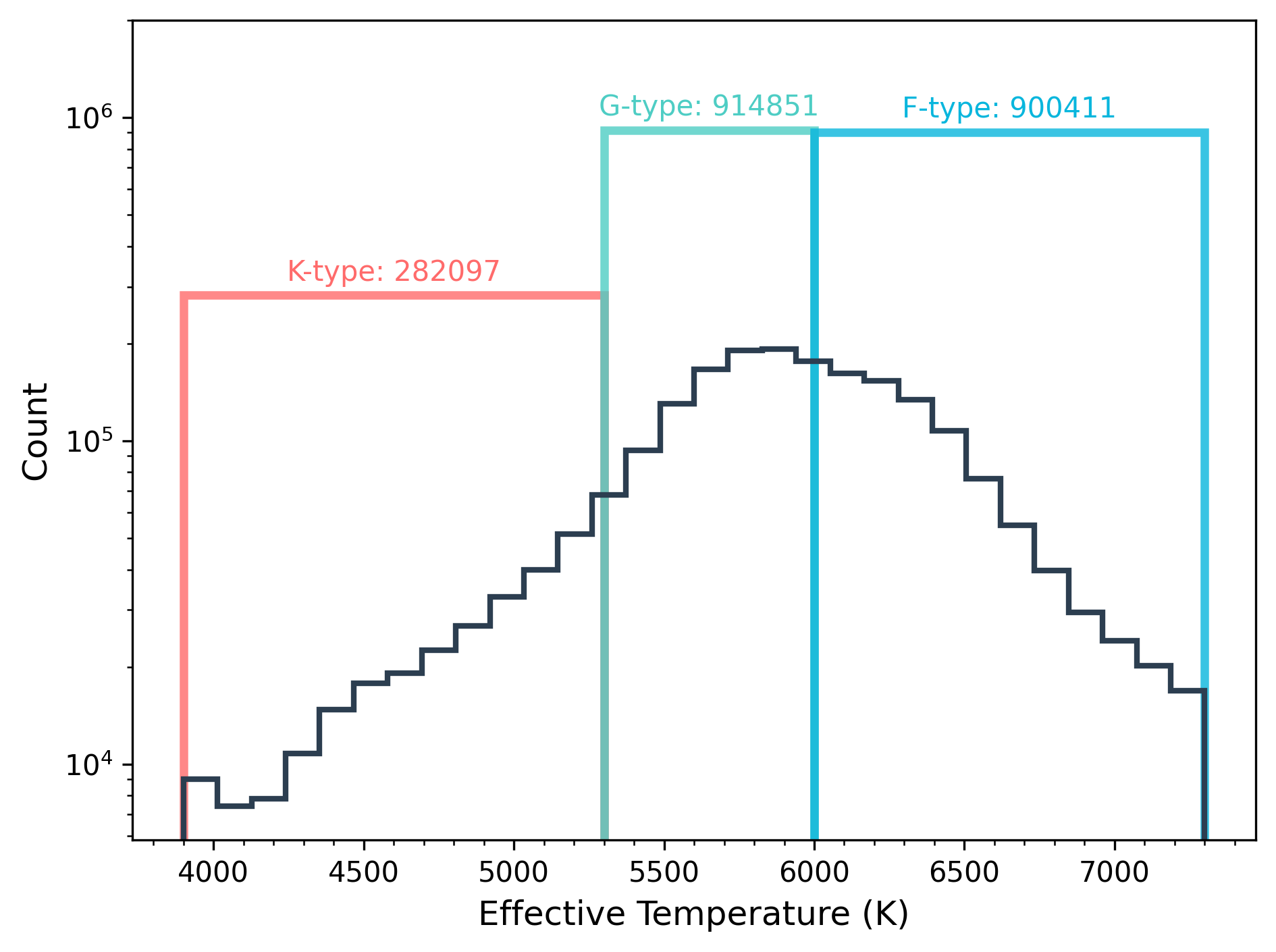}
    \caption{Effective temperature distribution of our \textit{TESS}-SPOC FGK main-sequence stellar sample. Boxes with different colours indicate the counts and temperature boundaries adopted for F-type ($6000$--$7300$\,K), G-type ($5300$--$6000$\,K), and K-type ($3900$--$5300$\,K) subsamples.}
    \label{fig:stellar-temperature}
\end{figure}

We subdivide this sample into three spectral-type subsamples based on TIC effective temperatures following the boundaries of \citet{kunimotoSearchingEntiretyKepler2020}:
\begin{itemize}
    \item F-type: $6000 < T_\mathrm{eff} \leq 7300$\,K,
    \item G-type: $5300 < T_\mathrm{eff} \leq 6000$\,K,
    \item K-type: $3900 \leq T_\mathrm{eff} \leq 5300$\,K.
\end{itemize}
{These divisions yield 900411 F-type stars, 914851 G-type stars, and 282097 K-type stars}. {Fig.~\ref{fig:stellar-temperature} shows the effective-temperature distribution of the parent sample, colour-coded by spectral-type subsample. The distribution is strongly non-uniform within each bin: because the sample is magnitude limited, cooler stars are only included out to smaller distances, so the K-type subsample is dominated by its hottest members (${\sim}5000$\,K) while the G- and F-type subsample is more evenly populated. We return to the consequences of this uneven sampling in Sections~\ref{subsec:hj-occ}.}

\subsection{Completeness Estimation}\label{subsec:completeness}

The completeness of our detection, vetting, and validation pipeline is estimated through injection-and-recovery tests on a simulated sample, as described in detail in Paper~I and \citet{hadjigeorghiouRAVENRAnkingValidation2026}. The key stages are: (i) observational bias (geometric transit probability and depth cut), (ii) detection efficiency from our GPU-based BLS pipeline, (iii) vetting completeness from the Non-simulated False Positive (NSFP) classifier, (iv) false positive (FP) completeness from the seven FP classifiers, and (v) signal-to-noise ratio (SNR) completeness from our trapezoidal-fit SNR threshold.

For this work, we recompute the completeness maps separately for each spectral-type subsample. Because stellar properties such as radius, luminosity, and noise characteristics differ across F, G, and K stars, the observational bias and detection efficiency vary accordingly, and the completeness cannot be reliably rescaled from the FGK result of Paper I. 
{For instance, stellar activity levels are different for different stellar types, making the different noises on the light curves. However, we do not need to treat them separately. In our work, simulated planetary systems around different stellar types always pass through the same end-to-end pipeline: BLS search, NSFP and FP classification, and the SNR cut. Any loss of sensitivity caused by differences in noise levels among stellar types is therefore incorporated into the spectral-type-specific completeness maps.}
We restrict the simulated stellar targets to the same effective temperature range as each subsample, ensuring that the completeness maps faithfully represent the sensitivity of our pipeline for each spectral type. Unlike {Paper~I}, we simulate the {\textit{Gaia}} RUWE values for the planetary systems with {\tt gaiamock}\footnote{\url{https://github.com/kareemelbadry/gaiamock}} \citep[][]{el-badryGenerativeModelGaia2024} based on their orbital parameters{, which allows the RUWE cut applied during vetting to be reproduced self-consistently in the injection-and-recovery tests}.

\begin{figure*}
    \includegraphics[width=\textwidth]{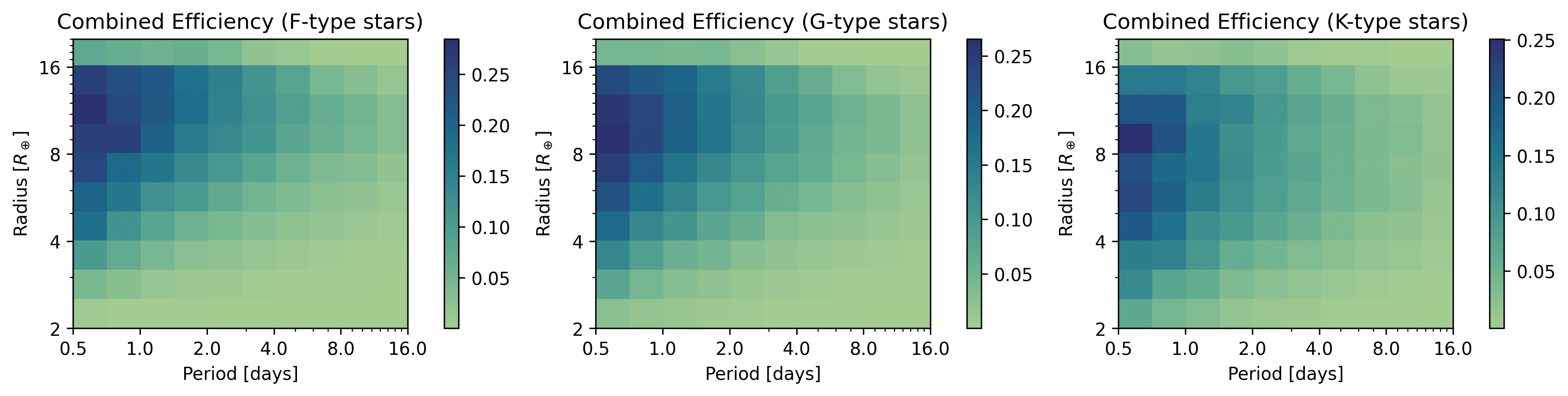}
    \caption{Combined completeness $Q(P, R_\mathrm{p})$ of our detection, vetting, and validation pipeline in the period--radius plane, shown for the F-type (left), G-type (middle), and K-type (right) subsamples. The maps are the product of the observational-bias, detection-efficiency, vetting, false-positive, and SNR completeness terms described in Section~\ref{subsec:completeness}, and provide the $Q(\boldsymbol{\omega})$ entering Equation~\ref{eq:hbm-likelihood}.}
    \label{fig:completeness}
\end{figure*}

The same thresholds as in Paper~I are adopted throughout: NSFP and FP classifier thresholds of 0.9, and an SNR threshold of 10. The non-uniform injection model is used as our fiducial model. We model completeness uncertainties using Beta distributions with Jeffreys priors {\citep[][]{jeffreysInvariantFormPrior1946}}, as in Paper~I. {Fig.~\ref{fig:completeness}} shows the combined completeness for F-, G-, and K-type stars, i.e.\ the end-to-end completeness of the five stages listed above. {In general, the completeness map follows the expected trend that planets with shorter orbital periods and larger radii are recovered more efficiently. The sharp decline in completeness above 16 $R_\oplus$ arises for two reasons. First, the simulated planets used to train our FP classifiers span only 1–16 $R_\oplus$, so signals implying larger radii fall outside the planetary training set and are therefore more likely to be classified as EBs by the model. Second, transiting companions larger than $\sim$16 $R_\oplus$ are more likely to be low-mass stars or brown dwarfs than planets. Because injected planets pass through the same classifiers, this effect is included in our completeness maps and accounted for in the occurrence rate corrections.}

\subsection{Occurrence Rate Estimation}\label{subsec:occ-method}

We apply the same two occurrence rate estimation methods as in Paper~I: (i) the inverse detection efficiency method (IDEM) with a weighted kernel density estimate (wKDE) for qualitative visualization, and (ii) the hierarchical Bayesian Poisson-process model for quantitative measurement.

The hierarchical Bayesian log-likelihood is
\begin{equation}\label{eq:hbm-likelihood}
\begin{aligned}
    \ln \mathcal{L}(\boldsymbol{\theta}) = & - \int Q(\boldsymbol{\omega}) \Gamma(\boldsymbol{\omega} | \boldsymbol{\theta}) \, \mathrm{d}\boldsymbol{\omega} \\
    & + \sum_{k=1}^{K} r_k\cdot\ln\left( \frac{1}{N_k} \sum_{n=1}^{N_k} \frac{Q(\boldsymbol{\omega}_k^{(n)})\Gamma(\boldsymbol{\omega}_k^{(n)} | \boldsymbol{\theta})}{p(\boldsymbol{\omega}_k^{(n)})}\right),
\end{aligned}
\end{equation}
where $\Gamma(\boldsymbol{\omega}|\boldsymbol{\theta})$ is the occurrence rate density, $Q(\boldsymbol{\omega})$ is the completeness, $r_k$ is the reliability of the $k$-th candidate, and $\boldsymbol{\omega}_k^{(n)}$ denotes the $n$-th posterior sample of the physical parameters. Full details of the notation and implementation, including the MCMC sampling with {\tt emcee} \citep{foreman-mackeyEmceeMCMCHammer2013} and the transit fitting with {\tt juliet} \citep{espinozaJulietVersatileModelling2019}, are provided in Paper~I.

For the spectral-type-resolved analyses, we apply the occurrence rate estimation independently to each stellar subsample, using the corresponding spectral-type-specific completeness maps and planet candidates.

\subsection{One-Dimensional Occurrence Rate Modelling}\label{subsec:1d-method}

To investigate the significance of the Neptunian ridge and savanna, we extract one-dimensional (1D) occurrence rate profiles as a function of orbital period within the radius range $6$--$8\,R_\oplus$, which encompasses the ridge region identified by \citet{castro-gonzalezMappingExoNeptunianLandscape2024}. We compare three models for the 1D occurrence rate density $\Gamma(P)$ in this radius strip.

\subsubsection{Non-parametric model}
We divide the period range ($0.5$--$16$\,d) into 10 equal bins in $\ln P$ and treat the occurrence rate density in each bin as a free parameter. This model has 10 free parameters and makes no assumption about the functional form of $\Gamma(P)$. {It is used only for visualization and as a sanity check on the parametric fits; because its parameters are independent by construction, it is not included in the model comparison of Section~\ref{subsubsec:model-comparison}.}

\subsubsection{Power-law model}
We model the occurrence rate density as a single power law:
\begin{equation}\label{eq:powerlaw}
    \Gamma(P) = A \left(\frac{P}{\mathrm{day}}\right)^{\alpha},
\end{equation}
where $A$ is the amplitude and $\alpha$ is the power-law index. This model predicts a smooth, monotonic trend in occurrence rate with period and has two free parameters. 

\subsubsection{Broken Power-law model}
To model the possible transition from the innermost desert to outer region, we use a broken power-law model:
\begin{equation}\label{eq:broken-powerlaw}
    \Gamma(P) =
    \begin{cases}
    A \left(\dfrac{P}{\mathrm{day}}\right)^{\alpha_1}, & P < P_b, \\[6pt]
    A \left(\dfrac{P_b}{\mathrm{day}}\right)^{\alpha_1}\left(\dfrac{P}{P_b}\right)^{\alpha_2}, & P \ge P_b,
    \end{cases}
\end{equation}
where $P_b$ is the break point, $\alpha_1$ and $\alpha_2$ are the power-law indices below and above the break, respectively{, and $A$ is the amplitude of the inner segment, i.e.\ the occurrence-rate density extrapolated to $P = 1$\,d. The outer segment carries no independent normalization: it is fixed by requiring continuity at $P_b$, which is why $A$ is referenced to 1\,d here and in Equations~\ref{eq:powerlaw} and \ref{eq:pl_lognormal} rather than to the break period}. This model has four free parameters.

\subsubsection{Power-law-plus-log-normal model}
To capture a potential localized excess (the ridge), we add a log-normal component on a power-law model:
\begin{equation}\label{eq:pl_lognormal}
    \Gamma(P) = A \left(\frac{P}{\mathrm{day}}\right)^{\alpha} + \frac{B}{P\,\sigma\sqrt{2\pi}} \exp\left(-\frac{(\ln P - \mu)^2}{2\sigma^2}\right),
\end{equation}
where $B$ is the amplitude of the log-normal component, $\mu$ is the ln-period location of the peak, and $\sigma$ is the width. This model has five free parameters.

\subsubsection{Model comparison}
For each model, we perform Bayesian MCMC sampling using the hierarchical Bayesian framework described in Section~\ref{subsec:occ-method}, adapted to the 1D case, using {\tt emcee} \citep{foreman-mackeyEmceeMCMCHammer2013}. For each parameter in the 1D models, we used broad, non-informative priors with ranges chosen to safely bracket all plausible values. Posterior sampling was performed with MCMC using 32 walkers. Convergence was assessed using the integrated autocorrelation time, \(\tau_{\mathrm{int}}\). We discarded the first \(50\,\tau_{\mathrm{int}}\) steps of each walker as burn-in, which is a conservative choice intended to ensure that the chains had fully forgotten their initial conditions. Only the post-burn-in samples were used for parameter estimation and model comparison. {We then compare the parametric models using} Pareto-smoothed importance-sampling leave-one-out (PSIS-LOO) cross-validation \citep{vehtariPracticalBayesianModel2017}, which estimates the expected log pointwise predictive density (ELPD). The difference in ELPD between models, $\Delta\mathrm{ELPD}$, divided by its standard error {$\mathrm{SE}_\Delta$} provides a measure of the significance with which one model is preferred over another. {We follow the {demonstration} of regarding $|\Delta\mathrm{ELPD}|/\mathrm{SE}_\Delta < 2$ as indicating that the data do not meaningfully distinguish between the two models \citep[][]{vehtariPracticalBayesianModel2017}. We emphasise that PSIS-LOO compares predictive performance rather than testing a hypothesis, so the quoted $\sigma$ values should not be read as detection significances.}

\subsection{Neptunian Desert Boundary Modelling}\label{subsec:boundary-method}
The boundaries of the Neptunian desert have been progressively refined as the number of known short-period planets has increased. \citet{szaboShortperiodCensorSubJupiter2011} first identified the ``sub-Jovian desert'' as a pronounced scarcity of intermediate-sized planets at very short orbital periods. \citet{mazehDearthShortperiodNeptunian2016} subsequently quantified its upper and lower boundaries in both the period--mass and period--radius planes, establishing the empirical limits that have been widely adopted in later studies. 

Using the completeness-corrected \textit{Kepler} population, \citet{castro-gonzalezMappingExoNeptunianLandscape2024} revised this picture by defining the desert statistically and identifying a distinct over-density of planets, termed the Neptunian ridge, between the desert and the more populated Neptunian savanna \citep[][]{bourrierDREAMOrbitalArchitecture2023}. More recently, \citet{osbornConfirmationHotSuperNeptune2026} investigated whether these boundaries vary with host-star spectral type. By comparing planets around FGK stars and M dwarfs, they found tentative evidence that the boundary occurs at shorter orbital periods around M dwarfs, although the difference is not statistically significant. Together, these studies show that the inferred desert boundaries depend on the available planet sample, the treatment of observational completeness, and potentially the properties of the host star.

{We therefore model the Neptunian desert in occurrence-rate space rather than in the distribution of detected planets. This has two advantages. First, the boundary of a completeness-corrected density is a statement about the underlying population, whereas the edge of a detected sample is a convolution of that population with the survey sensitivity, which varies rapidly across the desert. Second, a parametric description of the boundary can be compared directly and quantitatively with theoretical predictions, and its parameters can be propagated with their full posterior uncertainties. We describe the upper and lower boundaries separately below, because they are shaped by different physics and are best constrained by different parts of the plane.}

\subsubsection{Upper boundary}\label{subsubsec:upper-boundary}
{The upper boundary of the Neptunian desert is also the lower envelope of the hot Jupiter population, so a single parametrization constrains both.} We {model it} in the $\log P$--$\log R_\mathrm{p}$ plane following \citet{mazehDearthShortperiodNeptunian2016}, who characterized the desert edges as straight lines in the log--log plane. In contrast to their one-dimensional edge fit, we model the boundary in two dimensions across the entire upper-boundary radius range, $8 \le R_\mathrm{p} \le 16\,R_\oplus$, and infer it jointly with a smooth contrast in the occurrence-rate density across it{, so that the location of the boundary and the depth of the desert are constrained simultaneously rather than sequentially}.

Writing $u \equiv \log_{10}(P/\mathrm{day})$ and $v \equiv \log_{10}(R_\mathrm{p}/R_\oplus)$,
the boundary is set by a break radius {$R_b$} and two periods $(P_0, P_1)$: it is vertical at
constant period $P_0$ for $R_\mathrm{p} > R_b$, and a straight segment joining
$(P_0, R_b)$ to $(P_1,\,8\,R_\oplus)$ below{, so that $P_1 > P_0$ defines the outermost extent, or terminus, of the upper boundary}. Let $d(u,v)$ be the signed distance to this
boundary{, measured in the $(u,v)$ plane}, taken positive on the desert side. We turn it into a smooth gate with a logistic
sigmoid of width $\delta$ (in dex),
\begin{equation}
    G(u,v) = \left[\,1 + e^{-d(u,v)/\delta}\,\right]^{-1},
\end{equation}
and define the occurrence-rate density as a baseline log-density $\lambda$ suppressed by a
fractional depth $A \in (0,1)$ inside the desert,
\begin{equation}\label{eq:upper-boundary}
    {\Gamma}(P,R_\mathrm{p}) = 10^{\,\lambda\,[\,1 - A\,G(u,v)\,]}.
\end{equation}
The density therefore drops from $10^{\lambda}$ outside to $10^{\lambda(1-A)}$ on the
desert floor, a contrast of $\lambda A$ dex spread over a transition of width $\delta$.
The model has six free parameters, $(P_0, P_1, R_b, \lambda, A, \delta)$, all of which are inferred jointly within the hierarchical Bayesian framework of Section~\ref{subsec:occ-method}. The {red lines of} Fig.~\ref{fig:wKDE-F} shows the best-fit boundary model.
This two-segment model is motivated by the observed change in slope of the upper desert boundary, likely reflecting different physical processes. We defer discussion of the candidate mechanisms, and of what our spectral-type-resolved fits imply about them, to Section~\ref{subsec:discussion-desert}.

\subsubsection{Lower boundary}\label{subsubsec:lower-boundary}
The lower boundary of the desert is defined by the transition from the abundant sub-Neptune population to the depleted desert interior. Observationally, we find that, unlike the upper boundary, the occurrence rate across the lower desert region changes smoothly, without a sharp jump. We therefore model the sub-Neptune population using a two-dimensional power-law occurrence-rate density for \(2 \le R_\mathrm{p} \le 6\,R_\oplus\):
\begin{equation}\label{eq:2d-powerlaw}
    \Gamma(P, R_\mathrm{p}) = A \left(\frac{P}{\mathrm{day}}\right)^{\alpha} \left(\frac{R_\mathrm{p}}{R_\oplus}\right)^{\beta},
\end{equation}
where $A$ is the normalization{, and} $\alpha$ and $\beta$ are the period and radius power-law indices{, respectively}. We fit this model to the sub-Neptune region outside the desert using our Bayesian framework. The lower boundary is then defined as the contour in $P$--$R_\mathrm{p}$ space where the model-predicted occurrence rate density drops below a {chosen} threshold, corresponding to a transition into the depleted desert region. {We adopt a threshold of $\Gamma = 0.1$ (i.e.\ 0.1 planets per 100 stars per unit $\ln P\,\ln R_\mathrm{p}$, the units used throughout this paper and in the colour bars of Figs.~\ref{fig:wKDE-F}--\ref{fig:wKDE-K}), chosen so that the resulting contour connects continuously to the independently fitted upper boundary (Section~\ref{subsubsec:upper-boundary}).} We stress that, unlike the upper boundary, the lower boundary is therefore a threshold level set of a smooth transition rather than a feature inferred directly from the data, and its absolute position carries the systematic uncertainty of that threshold. Its relative variation between spectral types, which is what we use in Section~\ref{sec:discussion}, is insensitive to the threshold value because the same value is applied to all subsamples.

\section{Results}\label{sec:results}

\subsection{Planet Candidates}\label{subsec:planet-candidates}

Applying the same detection and \texttt{RAVEN} validation pipeline as in Paper~I to our FGK stellar sample, we identify {1301} planet candidates after alias removal, with periods ranging from 0.5 to 16\,d and radii from 2 to {20}\,$R_\oplus$. We subdivide these candidates by host star spectral type into the F (340), G (628), and K (333) subsamples. {The three subsamples therefore contain, respectively, 26, 48, and 26 per cent of the candidates, so the statistical precision achievable for the F- and K-type subsamples is roughly a factor of $\sqrt{2}$ worse than for the G-type subsample; this is the dominant limitation on the spectral-type comparisons that follow.} Full details of the detection pipeline and candidate selection are provided in Paper~I; the \texttt{juliet}-fitted transit parameters are taken from \citet{lafargaAutomaticSearchTransiting2026}.

\subsection{Occurrence Rates by Spectral Type}\label{subsec:spectral-occ}

\subsubsection{Weighted kernel density estimates}\label{subsubsec:wKDE}

\begin{figure}
    \centering
    \includegraphics[width=\columnwidth]{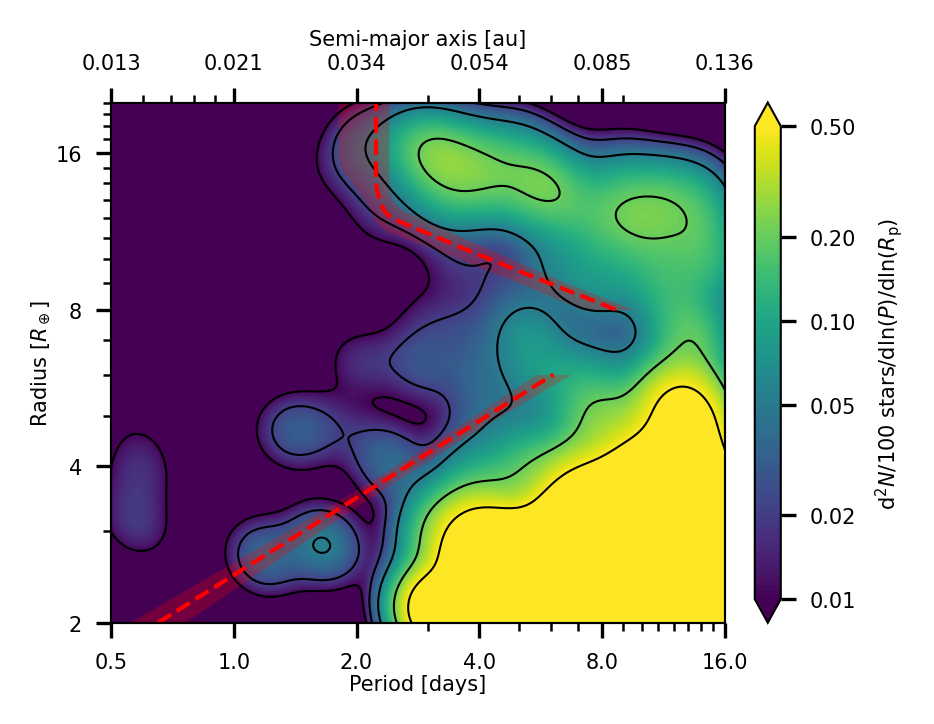}
    \caption{wKDE occurrence rate map for F-type stars in the period--radius plane. The colour scale is normalized logarithmically. The semi-major axis shown on the upper axis is a rough conversion based on the typical stellar mass of F-type stars, $1.3\,M_\odot$, and is indicative only. The dashed red lines show the modelled upper and lower boundaries based on occurrence rates from Section~\ref{subsec:desert-boundary}, with the shaded bands giving the $1\sigma$ credible intervals. Our KDE maps and desert boundary plots can be reproduced with
    {\tt NDesPlot} (\url{https://github.com/ckm3/NDesPlot}).}
    \label{fig:wKDE-F}
\end{figure}

\begin{figure}
    \centering
    \includegraphics[width=\columnwidth]{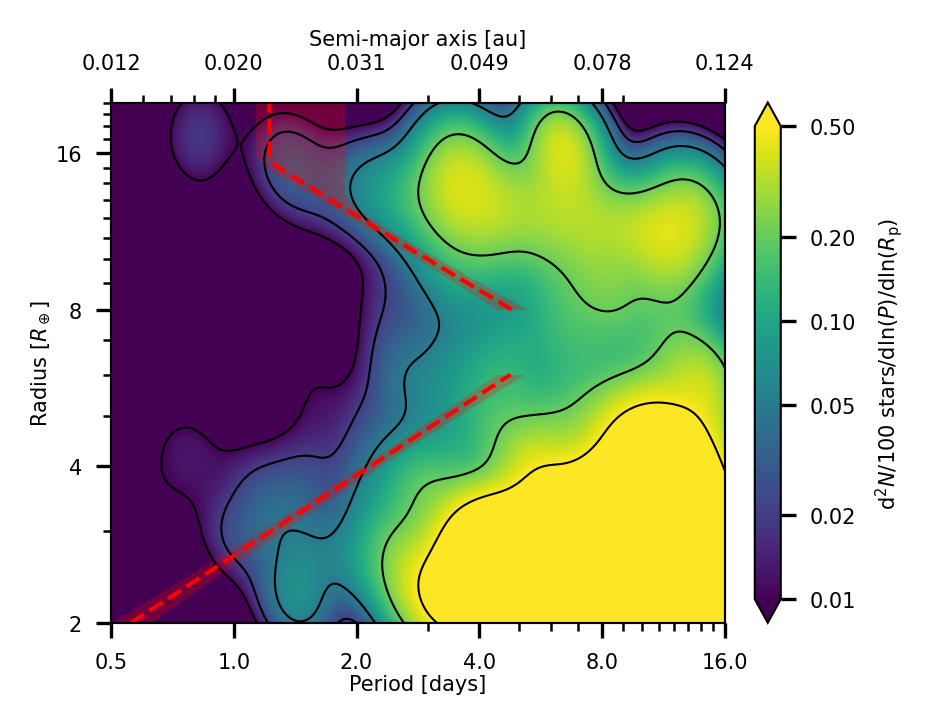}
    \caption{Same as Fig.~\ref{fig:wKDE-F} but for G-type stars. The semi-major axis is a rough estimation based on the typical stellar mass of {G-type} stars, 1.0\,$M_\odot$.}
    \label{fig:wKDE-G}
\end{figure}

\begin{figure}[!h]
    \centering
    \includegraphics[width=\columnwidth]{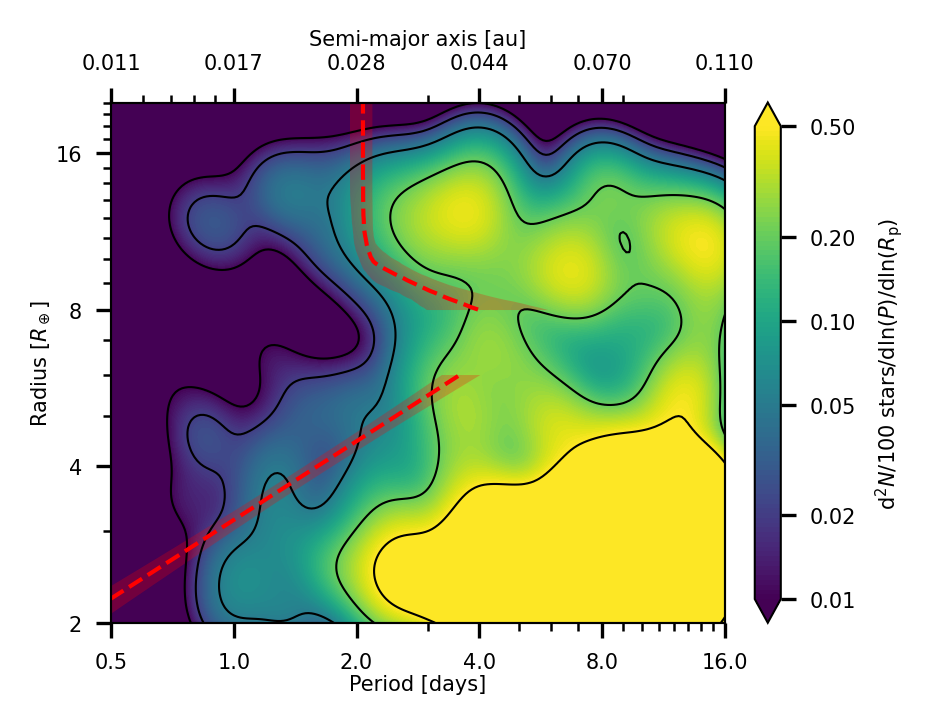}
    \caption{Same as Fig.~\ref{fig:wKDE-F} but for K-type stars. The semi-major axis is a rough estimation based on the typical stellar mass of {K-type} stars, 0.7\,$M_\odot$.}
    \label{fig:wKDE-K}
\end{figure}

We first examine the occurrence rate distributions qualitatively using the wKDE method (Section~\ref{subsec:occ-method}). Figs.~\ref{fig:wKDE-F}, \ref{fig:wKDE-G}, and \ref{fig:wKDE-K} show the wKDE occurrence rate maps for F-, G-, and K-type stars, respectively. {We stress that these maps are inverse-detection-efficiency estimates and are intended for visualization only; all quantitative statements below are based on the hierarchical Bayesian fits. Three features are common to all three subsamples.}
{First, the} abundance of sub-Neptunes dominates the overall occurrence{, rising towards longer periods across the whole radius range below ${\sim}4\,R_\oplus$. Second, the} hot Jupiter pile-up is present across all spectral types, appearing as a distinct island of enhanced density near $P \approx 3$--$4$\,d and $R_\mathrm{p} \gtrsim 10\,R_\oplus$ that is separated from the sub-Neptune population by the desert. Third, the Neptunian desert is visible in all three subsamples, though its extent and contrast vary: K-type hosts show the desert with the smallest extent, while F-type hosts exhibit a broader, less smoothly bounded depletion, with G-type hosts intermediate between the two.

\subsubsection{Binned two-dimensional occurrence rates}\label{subsubsec:2d-occ}

\begin{figure*}
    \centering
    \includegraphics[width=\textwidth]{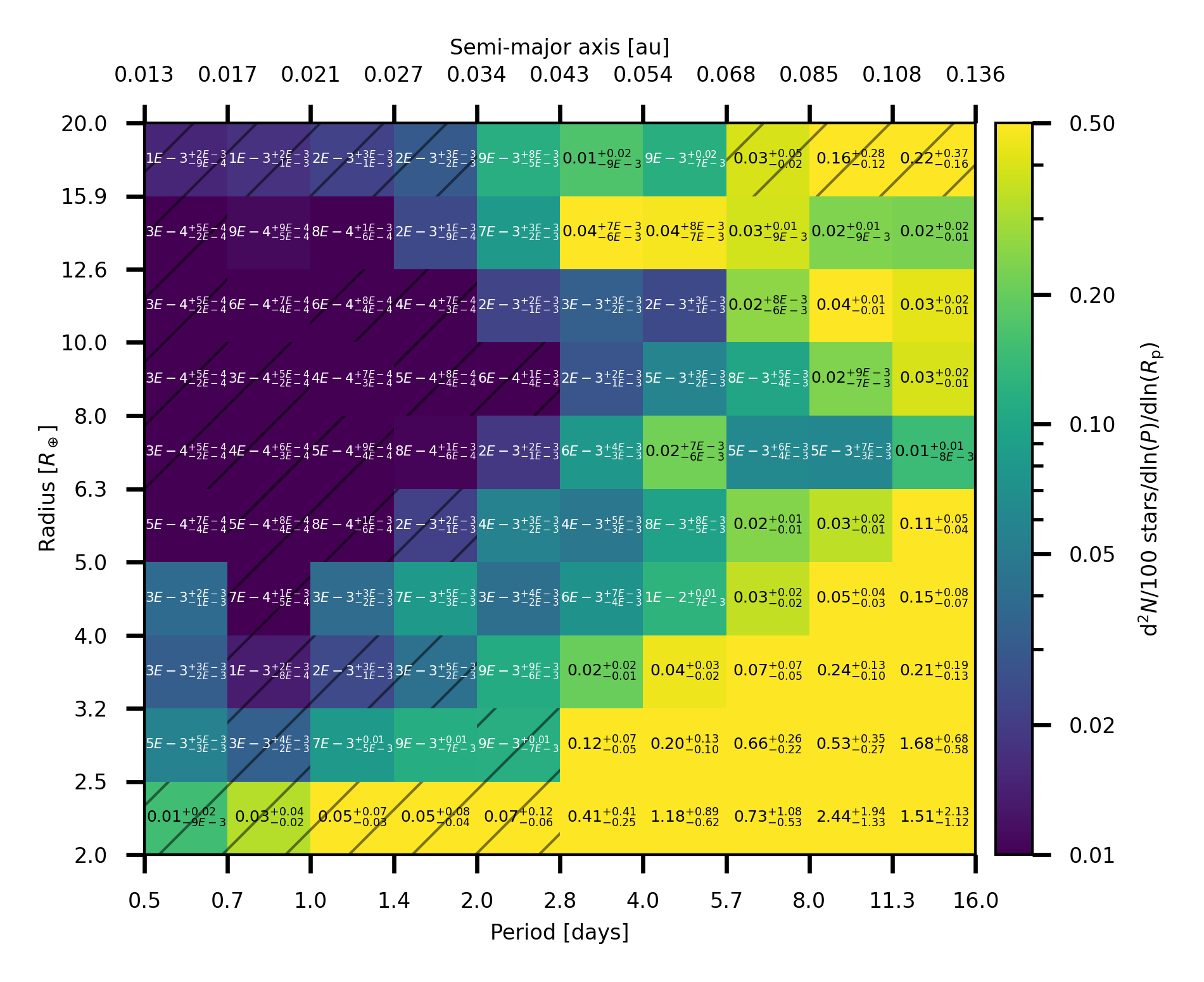}
    \caption{Occurrence rates of planet candidates around \textit{TESS}-SPOC F-type main-sequence stars on a $10\times10$ grid. Each cell displays the median occurrence rate as a percentage (planets per 100 stars) with uncertainties from the 16th and 84th percentiles. The colour encodes the occurrence rate density {$\Gamma$, in planets per 100 stars per unit $\ln P\,\ln R_\mathrm{p}$,} on a logarithmic scale{, truncated at $0.01$--$0.5$ to enhance visibility}. Bins with zero detections are overlaid with diagonal lines. We also provide an \href{https://cuikaiming.com/TESS-SPOC-Occurrence-Rate/}{online interactive plot} that allows users to easily select any region and obtain its total occurrence rate.}
    \label{fig:occ-rate-F}
\end{figure*}

\begin{figure*}
    \centering
    \includegraphics[width=\textwidth]{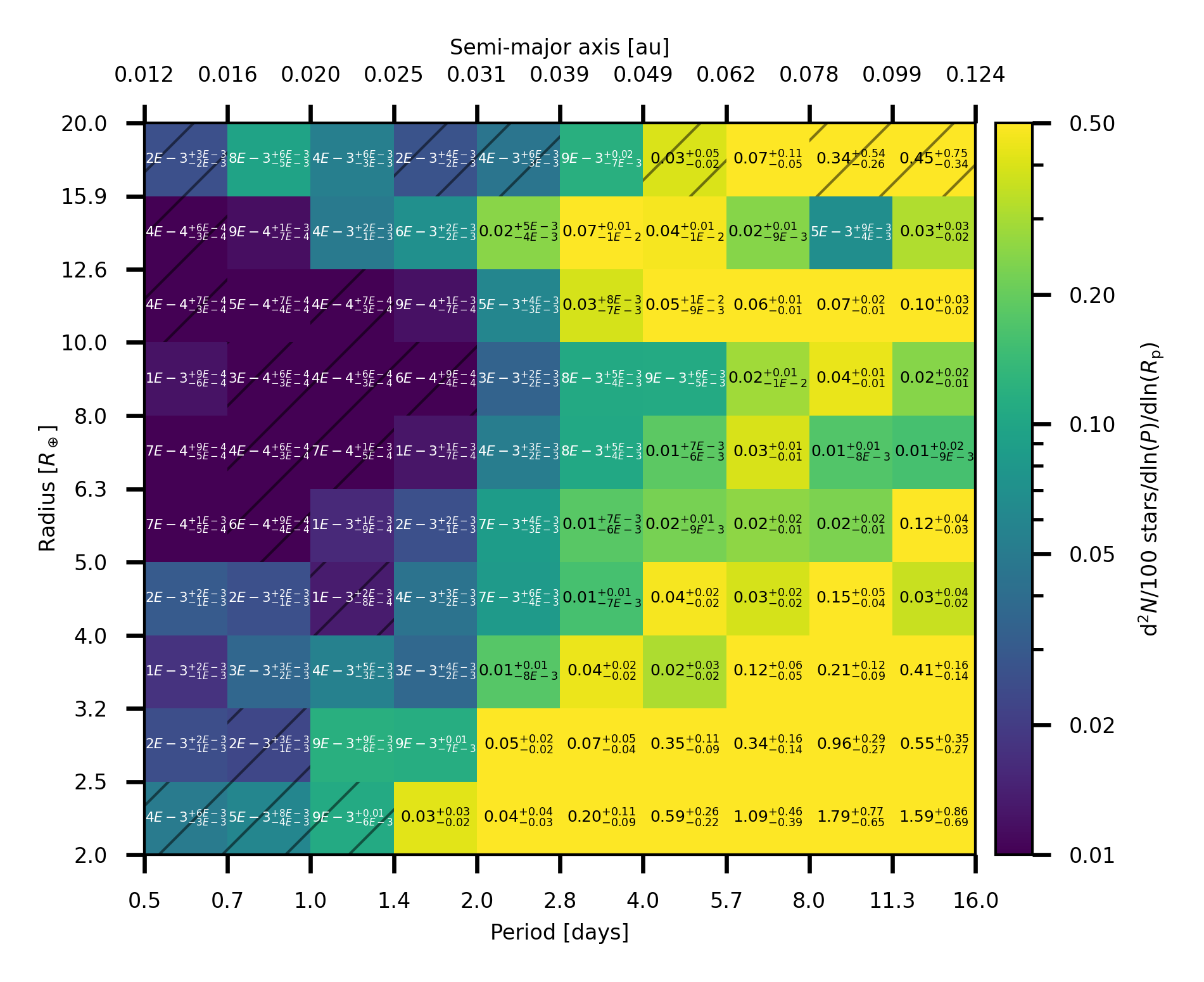}
    \caption{Same as Fig.~\ref{fig:occ-rate-F} but for G-type stars.}
    \label{fig:occ-rate-G}
\end{figure*}

\begin{figure*}
    \centering
    \includegraphics[width=\textwidth]{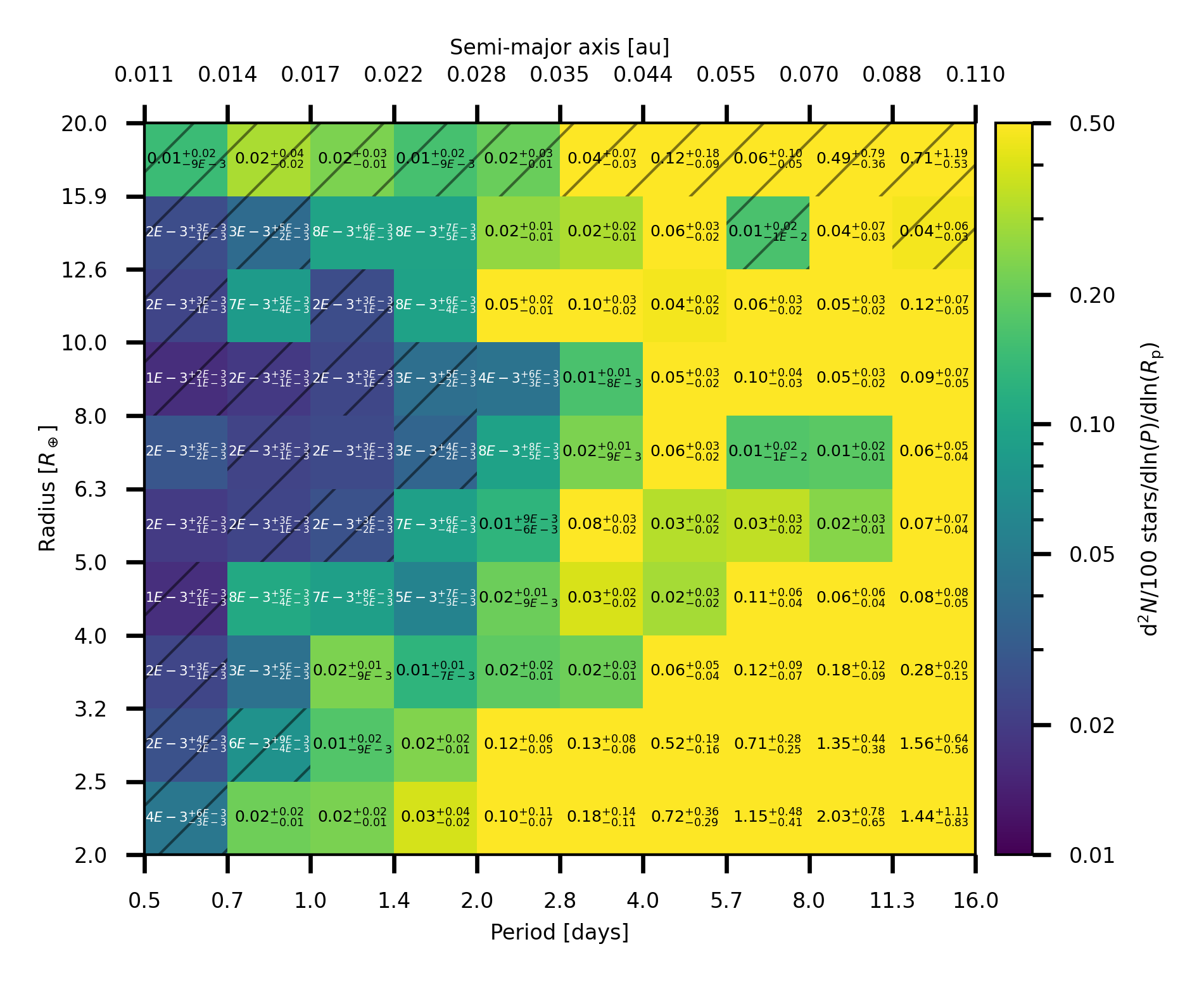}
    \caption{Same as Fig.~\ref{fig:occ-rate-F} but for K-type stars.}
    \label{fig:occ-rate-K}
\end{figure*}

We quantify the occurrence rates using our hierarchical Bayesian method on a $10\times10$ grid in $\log P$--$\log R_\mathrm{p}$ space, identically to Paper~I. Figs.~\ref{fig:occ-rate-F}, \ref{fig:occ-rate-G}, and \ref{fig:occ-rate-K} present the resulting occurrence rate maps for F-, G-, and K-type stars, respectively. The overall distribution is consistent with the wKDE maps. The value in each bin represents the average occurrence, expressed as a percentage (the number of planets per 100 stars of that spectral type), with the 16th and 84th percentiles indicating the lower and upper uncertainties, respectively. Appendix Tables~\ref{tab:TESS-SPOC 10x10 table F type}, \ref{tab:TESS-SPOC 10x10 table G type}, and \ref{tab:TESS-SPOC 10x10 table K type} list the results. {For practical use we also provide an online interactive version of these maps (\url{https://cuikaiming.com/TESS-SPOC-Occurrence-Rate/}), which allows an arbitrary region of the grid to be selected and its integrated occurrence rate returned directly.} We have also made the {full} posterior distributions of the occurrence rate density for each cell available online\footnote{\href{https://doi.org/10.5281/zenodo.21889834}{10.5281/zenodo.21889834}}, {so that the individual cells can be recombined into arbitrary period--radius regions without the independence assumption implied by the tabulated percentiles}.

Unlike the overall FGK sample, the F-, G-, and K-type subsamples contain fewer planets, leaving some bins with no detections. Estimating occurrence rates in these bins may introduce additional bias, especially where the completeness is low. Excluding bins with no detections, the overall occurrence rates within our parameter space ($0.5 < P < 16\,\mathrm{d}$, $2 < R_\mathrm{p} < 20\,R_\oplus$) are $12.1^{+3.0}_{-2.3}$, $10.2^{+1.2}_{-1.0}$, and $13.1^{+1.4}_{-1.2}$ {planets per 100 stars} for F-, G-, and K-type stars, respectively.

Two caveats attach to these integrated numbers. First, because the set of occupied bins differs between subsamples, the three values are integrals over slightly different regions of the period--radius plane and are therefore not strictly comparable; the F-type subsample, having fewer candidates, has the largest number of excluded bins. Second, and consequently, the ordering F $>$ G is not significant: the F-type value agrees with both the G-type ($0.7\sigma$) and the K-type ($0.3\sigma$) values, and only the K--G difference approaches significance (${\sim}1.7\sigma$). We therefore do not claim a monotonic increase in the integrated occurrence rate towards later spectral types. As Figs.~\ref{fig:occ-rate-F}--\ref{fig:occ-rate-K} show, the spectral-type dependence is instead localized: it is strongest for sub-Neptunes at $P \gtrsim 3$\,d, where the individual bins differ by factors of a few, and is absent or reversed at the shortest periods and largest radii, where the integrated value is dominated by the sparse, low-completeness bins.

\subsection{Neptunian Desert Boundaries}\label{subsec:desert-boundary}

\subsubsection{Upper boundary}\label{subsubsec:upperboundary}

\begin{figure}
    \centering
    \includegraphics[width=\linewidth]{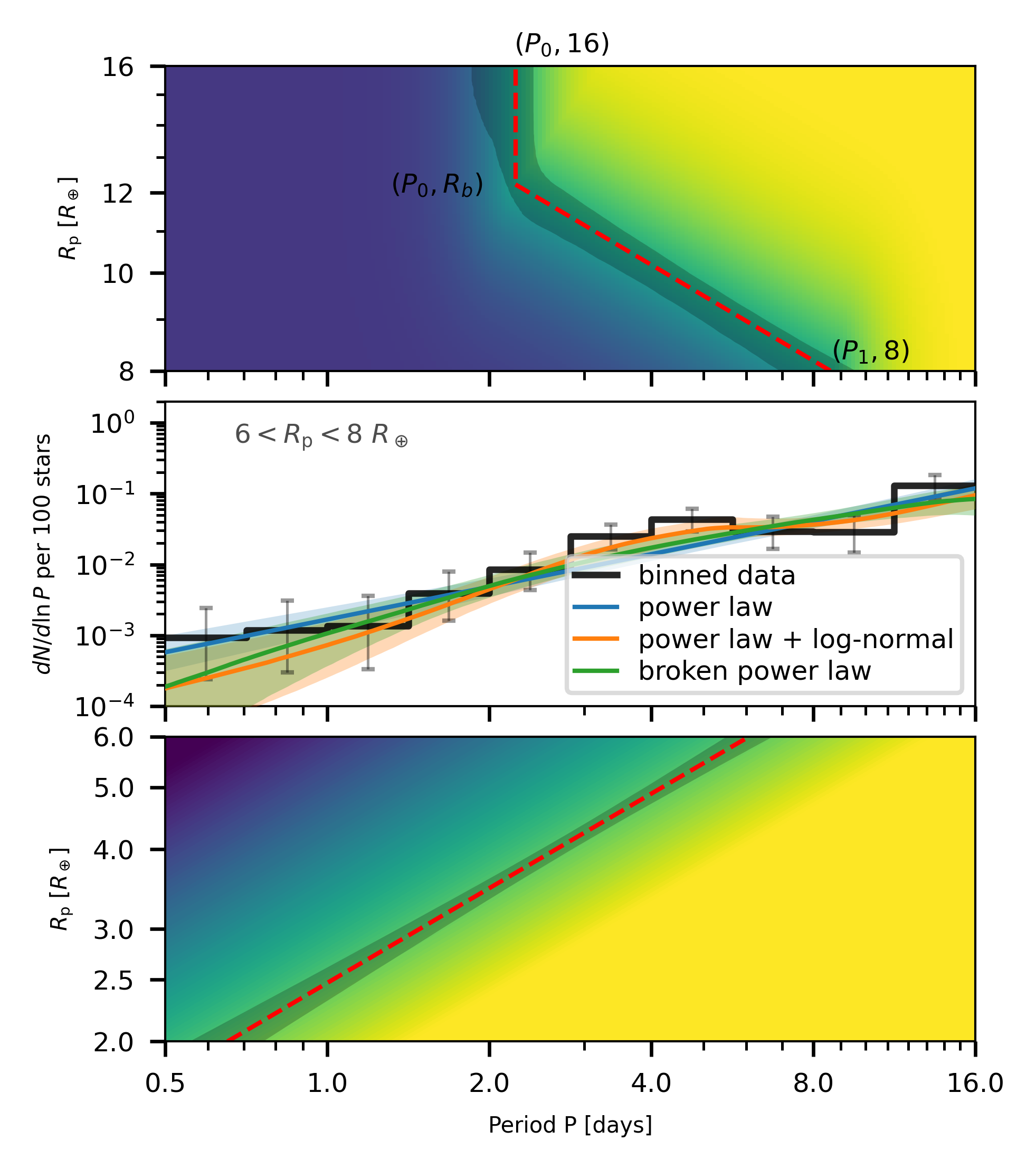}
    \caption{Occurrence-rate boundaries for F-type stars. Top panel: best-fitting model for the upper boundary (Section~\ref{subsubsec:upper-boundary}). The three coordinates of the upper boundary model are indicated. Middle panel: one-dimensional occurrence-rate density in the $6$--$8\,R_\oplus$ strip, shown together with three parametric models {(single power law, broken power law, and power law plus log-normal; Section~\ref{subsec:1d-method}) and the non-parametric binned estimate}. Bottom panel: best-fitting model for the lower boundary, overlaid with the 0.1 per cent occurrence-rate density contour. Shaded regions denote $1\sigma$ uncertainty bands.}
    \label{fig:F-type-boundaries}
\end{figure}

\begin{figure}
    \centering
    \includegraphics[width=\linewidth]{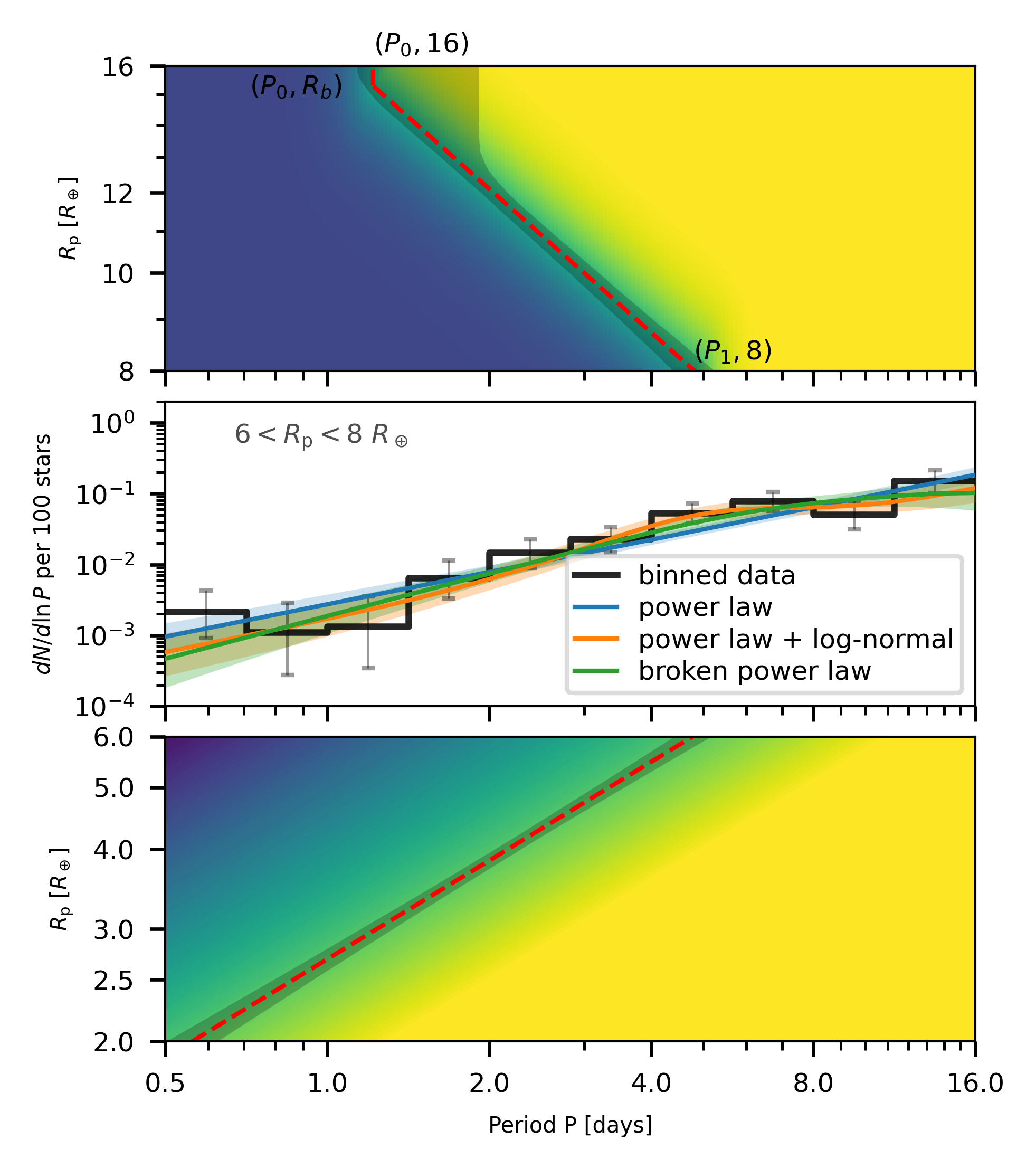}
    \caption{{Same as Fig.}~\ref{fig:F-type-boundaries}, but for G-type stars.}
    \label{fig:G-type-boundaries}
\end{figure}

\begin{figure}
    \centering
    \includegraphics[width=\linewidth]{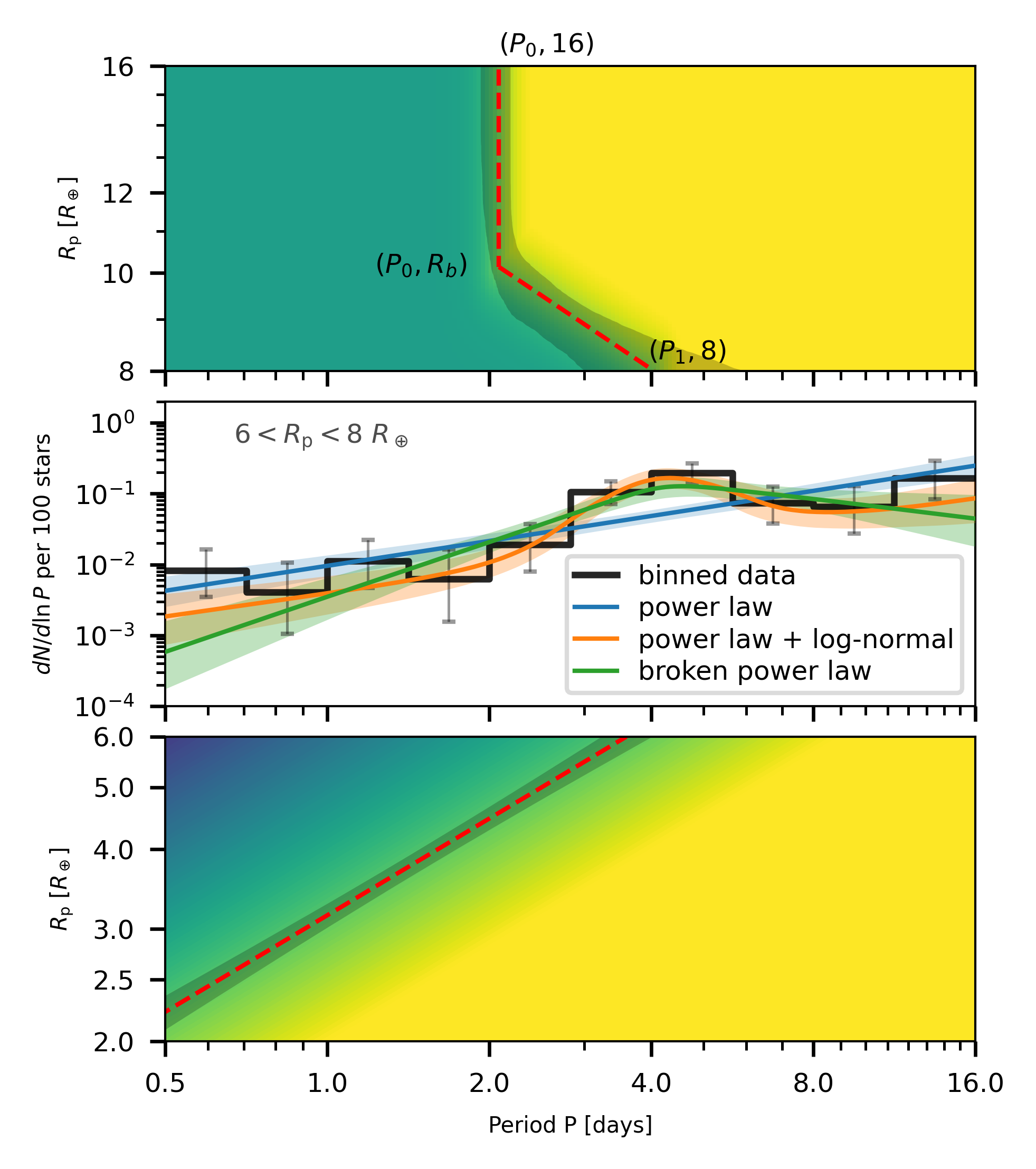}
    \caption{{Same as Fig.}~\ref{fig:F-type-boundaries}, but for K-type stars.}
    \label{fig:K-type-boundaries}
\end{figure}

We model the upper boundary of the Neptunian desert for each spectral type and for the combined FGK sample using the two-segment {model} described in Section~\ref{subsubsec:upper-boundary}. The results are shown in the upper panels of Figs.~\ref{fig:F-type-boundaries}, \ref{fig:G-type-boundaries}, and \ref{fig:K-type-boundaries}, and the corresponding parameters are summarized in Table~\ref{tab:upper-boundary}. In all four samples the fitted transition width is small, $\delta = 0.02$--$0.06$\,dex; since the logistic gate completes ${\sim}80$ per cent of its drop over an interval $4.4\delta$, the full contrast is established within a factor of ${\lesssim}2$ in orbital period. The upper boundary is therefore genuinely sharp rather than a gradual decline, which is the key requirement for interpreting it as a physical edge rather than as the tail of a smooth distribution. The inferred contrast across the boundary, $\lambda A$, is $2.28^{+0.37}_{-0.26}$\,dex for F-type, $2.14^{+0.37}_{-0.48}$\,dex for G-type, and $1.29^{+0.23}_{-0.20}$\,dex for K-type hosts, so the desert floor is roughly an order of magnitude less depleted {($2.9\sigma$)}, relative to its surroundings, around K-type stars than around F-type stars. The two-segment geometry is required by the data: the short-period segment is nearly vertical, consistent with Roche-lobe overflow or tidal disruption shaping the boundary at the shortest periods.

Comparing across spectral types, the two segments behave differently, and this contrast is the central result of this subsection. The near-vertical short-period edge sits at essentially the same orbital period in all three subsamples, $P_0 = 2.24^{+0.17}_{-0.39}$\,d (F), $1.22^{+0.67}_{-0.09}$\,d (G), and $2.08^{+0.11}_{-0.16}$\,d (K), with the F and K values statistically indistinguishable; the lower G-type value is driven by a markedly asymmetric posterior and we regard it as the least secure of the three. By contrast, the terminus of the boundary, $P_1$, moves inward monotonically from $8.62^{+0.95}_{-1.57}$\,d (F) through $4.81^{+0.42}_{-0.39}$\,d (G) to $4.05^{+2.00}_{-1.04}$\,d (K). The F--G difference is significant at $2.3\sigma$ and the F--K difference at $1.8\sigma$, so the trend is suggestive but not yet decisive; the G--K difference is not significant. Taken at face value, the outermost extent of the upper boundary lies at shorter orbital periods around cooler, lower-mass stars while its innermost edge does not move, which points to two distinct mechanisms operating at the two ends of the boundary. We discuss the interpretation of the two segments in Section~\ref{subsec:discussion-desert}.

The break radius $R_b$ is formally largest for G-type hosts ($15.29^{+0.46}_{-2.14}\,R_\oplus$) and smallest for K-type hosts ($10.13^{+1.09}_{-0.85}\,R_\oplus$), but the G-type value is pushed against the upper edge of the fitted radius range and its posterior is truncated, so we do not interpret the ordering as physical. Restricting attention to the F- and K-type fits, which are well within the fitted range, $R_b$ decreases towards cooler hosts at ${\sim}1.6\sigma$, in the sense expected if the vertical segment tracks a fixed planet density and the boundary radius scales with the maximum stable giant-planet size.

\begin{figure}
    \centering
    \includegraphics[width=1\linewidth]{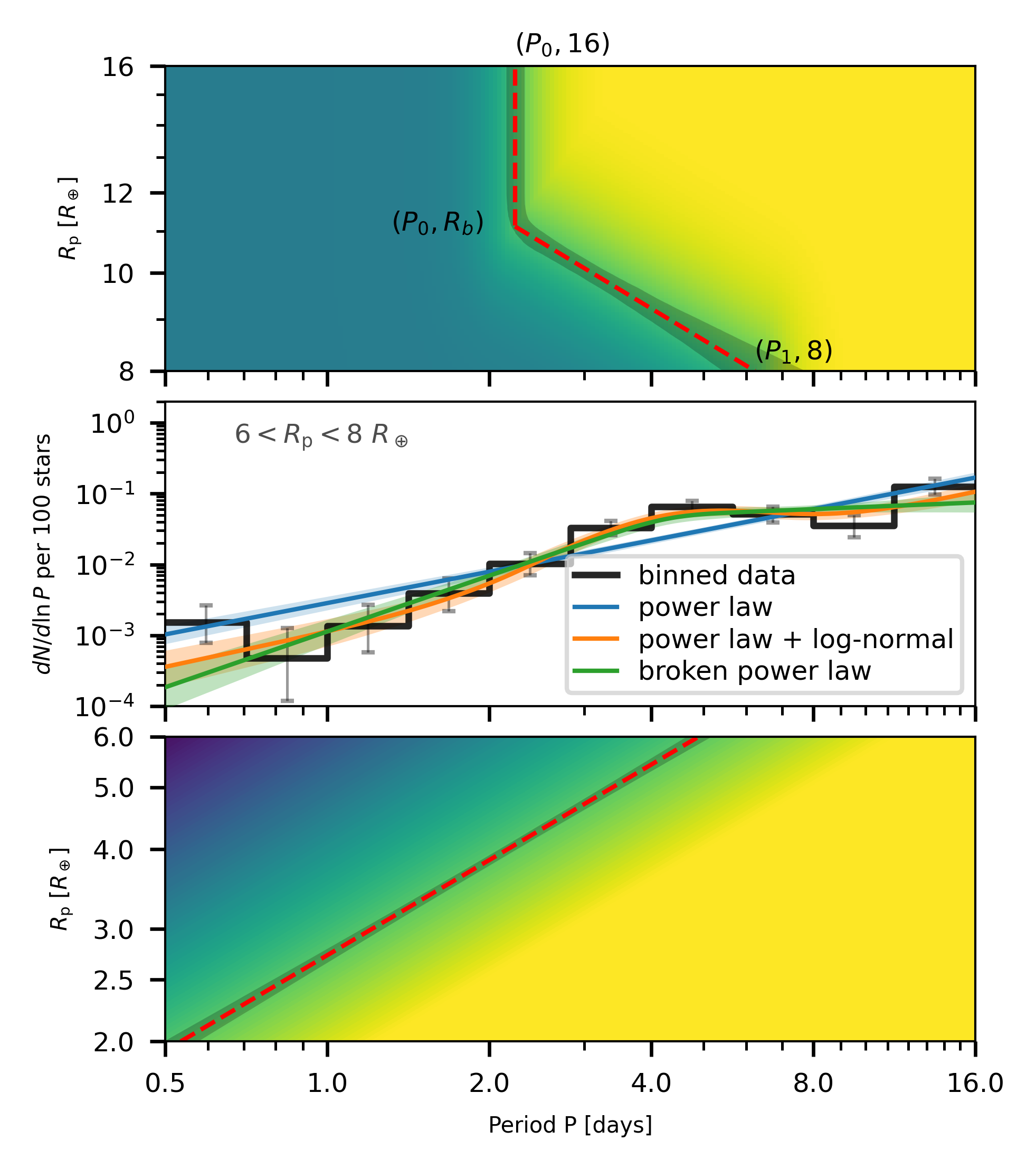}
    \caption{{Same as Fig.}~\ref{fig:F-type-boundaries}, but for {the combined FGK sample, shown for reference}.}
    \label{fig:FGK-boundaries}
\end{figure}

\begin{table*}
\centering
\caption{Best-fit parameters for the two-segment upper boundary model {of Section~\ref{subsubsec:upper-boundary}. $P_0$ is the period of the near-vertical segment, $P_1$ the period at which the boundary reaches $8\,R_\oplus$, $R_b$ the break radius, $\lambda$ the baseline log-density outside the desert, $A$ the fractional suppression inside it, and $\delta$ the logistic transition width in dex. As in Table~\ref{tab:lower-boundary}, $\lambda$ refers to the raw occurrence-rate density summed over the subsample, so $10^{\lambda}$ must be divided by $N_\star$ to obtain a per-star density; $A$ and $\delta$ are dimensionless and are directly comparable between rows. Values are medians with 16th--84th percentile uncertainties.}}
\label{tab:upper-boundary}
\renewcommand{\arraystretch}{1.3}
\setlength{\tabcolsep}{5pt}
\begin{tabular}{lcccccc}
\hline
Sample & $P_0$ (d) & $P_1$ (d) & $R_b$ ($R_\oplus$) & $\lambda$ & $A$ & $\delta$ \\
\hline
F   & $2.24^{+0.17}_{-0.39}$ & $8.62^{+0.95}_{-1.57}$ & $12.23^{+1.42}_{-0.75}$ & $3.68^{+0.11}_{-0.09}$ & $0.62^{+0.10}_{-0.07}$ & $0.06^{+0.02}_{-0.01}$ \\
G   & $1.22^{+0.67}_{-0.09}$ & $4.81^{+0.42}_{-0.39}$ & $15.29^{+0.46}_{-2.14}$ & $3.69^{+0.04}_{-0.03}$ & $0.58^{+0.10}_{-0.13}$ & $0.03 \pm 0.01$ \\
K   & $2.08^{+0.11}_{-0.16}$ & $4.05^{+2.00}_{-1.04}$ & $10.13^{+1.09}_{-0.85}$ & $3.23 \pm 0.06$ & $0.40^{+0.07}_{-0.06}$ & $0.02^{+0.02}_{-0.01}$ \\
FGK & $2.23 \pm 0.09$        & $6.23^{+1.62}_{-0.77}$ & $11.11^{+0.37}_{-0.35}$ & $4.04^{+0.05}_{-0.04}$ & $0.39 \pm 0.02$        & $0.04 \pm 0.01$ \\
\hline
\end{tabular}
\end{table*}

\begin{table}
\centering
\caption{Best-fit parameters for the 2D power-law lower boundary model {(Equation~\ref{eq:2d-powerlaw}), fitted to the sub-Neptune population over $2 \le R_\mathrm{p} \le 6\,R_\oplus$, where $\alpha$ and $\beta$ are the period and radius indices. $A$ is the normalization of the raw occurrence-rate density, i.e.\ planets per unit $\ln P\,\ln R_\mathrm{p}$ summed over the whole subsample rather than per star. It must therefore be divided by the number of stars $N_\star$ in that subsample (Section~\ref{subsec:sample}) to obtain a per-star occurrence-rate density, and the amplitudes are consequently {not} directly comparable between rows: the K-type subsample contains roughly three times fewer stars than the F- and G-type ones, so a similar $A$ implies a substantially higher per-star rate.}}
\label{tab:lower-boundary}
\renewcommand{\arraystretch}{1.3}
\begin{tabular}{lccc}
\hline
Sample & {$A/10^{4}$} & $\alpha$ & $\beta$ \\
\hline
F   & {$3.83^{+2.41}_{-1.48}$} & $2.04\pm0.11$ & $-4.15^{+0.35}_{-0.36}$ \\
G   & {$3.60^{+1.25}_{-0.94}$} & $1.90^{+0.08}_{-0.07}$ & $-3.71\pm0.23$ \\
K   & {$1.61^{+0.63}_{-0.45}$} & $1.77\pm0.08$ & $-3.53\pm0.27$ \\
FGK & {$9.98^{+2.23}_{-1.83}$} & $1.91\pm0.05$ & $-3.85\pm0.16$ \\
\hline
\end{tabular}
\end{table}

\subsubsection{Lower boundary}

The lower boundary of the Neptunian desert is defined by the transition from the well-populated sub-Neptune regime to the depleted desert. Following the method described in Section~\ref{subsubsec:lower-boundary}, we fit a 2D power-law model (Equation~\ref{eq:2d-powerlaw}) to the sub-Neptune occurrence rates and then define the lower boundary as the contour at which the model-predicted occurrence rate density equals a manually chosen threshold value.

The lower panels of Figs.~\ref{fig:F-type-boundaries}, \ref{fig:G-type-boundaries}, and \ref{fig:K-type-boundaries} show the resulting lower boundaries for each spectral type, and {Table~\ref{tab:lower-boundary}} lists the corresponding parameters. The 2D power-law model generally captures the abundance distribution of the sub-Neptune population across all stellar types. We adopt the 0.1 per cent occurrence-rate density contour as the boundary, {because it is the level set that joins smoothly onto the independently fitted upper boundary at $R_\mathrm{p} \approx 6$--$8\,R_\oplus$; adopting a threshold a factor of two higher or lower translates the contour by ${\lesssim}0.1$\,dex in radius without altering the relative ordering of the three spectral types}.

The lower boundary shows a mild spectral-type dependence, driven by a systematic change in both fitted indices from F- to K-type hosts. The period index decreases from $\alpha = 2.04 \pm 0.11$ (F) through $1.90^{+0.08}_{-0.07}$ (G) to $1.77 \pm 0.08$ (K), a total change significant at $2.0\sigma$, while the radius index becomes shallower, from $\beta = -4.15^{+0.35}_{-0.36}$ (F) to $-3.53 \pm 0.27$ (K), a change of $1.4\sigma$. Both indices thus vary in the same direction, and their combined effect is to move the 0.1 per cent contour upward in radius and inward in period as the host star cools, so that the desert becomes less extended towards small radii and short periods around K-type stars. Equivalently, the sub-Neptune population encroaches further into the desert from below around cooler hosts, the same conclusion reached independently from the upper-boundary contrast $\lambda A$, which is lowest for K-type hosts. That two independently fitted boundaries agree on the direction of the trend is reassuring, since they use disjoint radius ranges ($2$--$6\,R_\oplus$ and $8$--$16\,R_\oplus$) and therefore disjoint sets of planet candidates.
Overall, the combined picture shows a desert which grows smaller from both the upper and lower boundaries for cooler host stars.

\subsection{The Neptunian Ridge and Savanna}\label{subsec:ridge-savanna}

\subsubsection{One-dimensional occurrence rate profiles}

To investigate the Neptunian ridge and savanna, we extract the 1D occurrence rate density as a function of orbital period for planets in the $6$--$8\,R_\oplus$ radius range. This radius strip was chosen because it lies along the outer edge of the Neptunian desert, where the ridge{, a localized enhancement in occurrence rate at periods of ${\sim}3$--$6$\,d,} has been previously reported \citep[][]{castro-gonzalezMappingExoNeptunianLandscape2024}.

We present the analysis in the middle panels of Figs.~\ref{fig:F-type-boundaries}, \ref{fig:G-type-boundaries}, and \ref{fig:K-type-boundaries}. The non-parametric occurrence rate density {estimates show} that the overall increase from 0.5\,d to 16\,d is generally well captured by a power-law model{, with indices $\alpha = 1.54^{+0.23}_{-0.22}$ (F), $1.52^{+0.19}_{-0.18}$ (G), and $1.17^{+0.22}_{-0.21}$ (K) that are consistent with one another}. Beyond 4--6\,d, however, this increase appears to level off. The higher occurrence-rate bin near 16\,d lies too close to the edge of our sample to constrain the slope reliably{, since it is bounded on one side by the period limit of the survey}. We also observe a {comparatively pronounced} bump and dip in the K-type occurrence rate {distribution}, which {is qualitatively} consistent with the ``ridge'' and ``savanna''.

To test the significance of the local excess and deficit in the occurrence-rate estimates, we consider three different parametric models. A single power-law model {captures} the global trend {and serves as the null hypothesis}. The second model, a power-law plus log-normal model, is intended to fit the ridge and savanna simultaneously. The third model is a broken power-law model, designed to compete with the power-law plus log-normal model and to test whether the local bump and dip are significant enough to warrant that interpretation, or whether they can instead be modelled as a plateau. {This third model is important because the two hypotheses make very different physical claims: a broken power law implies only that the occurrence rate stops rising beyond a characteristic period, whereas a log-normal excess implies a genuinely separate population superimposed on a smooth background. Distinguishing them is therefore the crux of the ridge question, and it is not addressed by comparing against a single power law alone.} {Table~\ref{tab:1d-parametric-models}} lists their best-fit parameters and corresponding uncertainties.

\begin{table*}
\centering
\caption{Best-fit parameters for the 1D {occurrence-rate} models {fitted to the $6$--$8\,R_\oplus$ radius strip (Section~\ref{subsec:1d-method}). The amplitudes $A$ and $B$ are raw occurrence-rate densities evaluated at $P = 1$\,d, in planets per unit $\ln P\,\ln R_\mathrm{p}$ summed over the whole subsample; as in Table~\ref{tab:lower-boundary}, they must be divided by $N_\star$ before being compared between rows. $\mu$ and $\sigma$ are in $\ln(P/\mathrm{day})$. Values are medians with 16th--84th percentile uncertainties.}}
\label{tab:1d-parametric-models}
\renewcommand{\arraystretch}{1.3}
\setlength{\tabcolsep}{2.6pt}
\begin{tabular}{lccccccccccc}
\hline
& \multicolumn{2}{c}{Power law}
& \multicolumn{4}{c}{Broken power law}
& \multicolumn{5}{c}{Power law + log-normal} \\
\cmidrule(lr){2-3}
\cmidrule(lr){4-7}
\cmidrule(lr){8-12}
Sample
& $\alpha$ & $A$
& $\alpha_1$ & $\alpha_2$ & $P_\mathrm{b}$ (d) & $A$
& $\alpha$ & $A$ &  $\mu$ & $\sigma$ & $B$ \\
\hline
F
& $1.54^{+0.23}_{-0.22}$
& $52.41^{+26.11}_{-19.03}$
& {$2.36^{+2.66}_{-0.83}$}
& {$0.88^{+0.66}_{-1.46}$}
& {$4.38^{+8.74}_{-2.58}$}
& {$34.13^{+38.24}_{-22.50}$}
& $1.81^{+0.53}_{-0.36}$
& $19.03^{+25.56}_{-14.01}$
& $1.79^{+0.20}_{-0.19}$
& $0.42^{+0.11}_{-0.17}$
& $1461.20^{+1103.68}_{-806.09}$ \\
G   & $1.52^{+0.19}_{-0.18}$
& $87.22^{+33.36}_{-25.71}$
& $1.99^{+0.57}_{-0.42}$
& $0.29^{+0.88}_{-1.41}$
& $6.48^{+5.49}_{-2.39}$
& $59.70^{+35.42}_{-26.11}$
& $1.54^{+0.27}_{-0.26}$
& $52.79^{+32.53}_{-24.14}$
& $1.91^{+0.12}_{-0.18}$
& $0.38^{+0.11}_{-0.12}$
& $2642.33^{+1775.85}_{-1417.84}$ \\
K
& $1.17^{+0.22}_{-0.21}$
& $94.07^{+38.34}_{-29.90}$
& {$2.58^{+0.69}_{-0.60}$}
& {$-0.93^{+0.77}_{-0.90}$}
& {$4.24^{+0.72}_{-0.48}$}
& {$34.37^{+29.20}_{-18.03}$}
& $1.10\pm0.37$
& $38.10^{+29.00}_{-19.05}$
& $1.53^{+0.17}_{-0.09}$
& $0.24^{+0.10}_{-0.06}$
& $1706.58^{+816.51}_{-560.88}$ \\
FGK
& $1.47\pm0.12$
& $208.10^{+49.50}_{-42.43}$
& {$2.61^{+0.43}_{-0.36}$}
& {$0.31^{+0.34}_{-0.38}$}
& {$4.27^{+0.69}_{-0.61}$}
& {$82.63^{+38.57}_{-29.13}$}
& $1.64^{+0.22}_{-0.19}$
& $81.31^{+42.14}_{-33.03}$
& $1.73^{+0.17}_{-0.12}$
& $0.36^{+0.09}_{-0.08}$
& $5913.66^{+2553.90}_{-1813.81}$ \\
\hline
\end{tabular}
\end{table*}

\subsubsection{Model comparison}\label{subsubsec:model-comparison}

We compare the three 1D models using PSIS-LOO cross-validation for each spectral type and for the combined FGK sample. The results are summarized in Table~\ref{tab:loo}{, where $\Delta\mathrm{ELPD}$ is quoted relative to the best-performing model in each row, so that the preferred model has $\Delta\mathrm{ELPD} = 0$ by construction}. In {three of the four samples the power-law-plus-log-normal model is the preferred model}, suggesting that including a log-normal component can improve the fit. However, the improvement is generally modest{, and in no case does it reach the $|\Delta\mathrm{ELPD}|/\mathrm{SE}_\Delta > 2$ level conventionally required to prefer one model over another}. For the combined FGK sample, the difference between the broken power-law and the power-law-plus-log-normal model is only about $1.4\sigma$ when measured relative to its standard error, and even the preference over the single power-law model is only marginal at $2.3\sigma$. {For the G-type subsample the ordering actually reverses, with the broken power law {slightly} preferred ($0.1\sigma$), which is another way of saying that the two models are indistinguishable.} We also tested different radius ranges, such as $5.5$--$8.5\,R_\oplus$; among them, the $6$--$8\,R_\oplus$ range shows the highest significance. {We note that this choice was made after inspecting the data, so the quoted significances should be regarded as upper limits on the true preference for a ridge component.}

The indication of an additional ridge-like component is somewhat more pronounced for K-type stars. In particular, the middle panel of Fig.~\ref{fig:K-type-boundaries} shows a noticeable dip around 8\,d, and the power-law-plus-log-normal model is favoured over the single power law at the $1.6\sigma$ level and over the broken power law at the $1.4\sigma$ level. However, the overall model comparison still does not provide strong statistical evidence for such a feature. Therefore, rather than claiming a significant detection of a ridge, we conclude that its existence remains uncertain and that the current data do not clearly require a distinct ridge component. Our measurements show that a completeness-corrected, homogeneously selected \textit{TESS} sample, restricted to $P < 16$\,d and $6$--$8\,R_\oplus$, can be equally well described by smooth alternative models without invoking a separate ridge. Distinguishing between these scenarios will require either a substantially larger candidate sample in this radius range or additional constraints beyond occurrence rates, such as eccentricities and stellar obliquities \citep[][]{bourrierDREAMOrbitalArchitecture2023, hallattRevealingOriginDesert2026}.

\begin{table*}
\centering
\caption{LOO cross-validation model comparison results for the 1D occurrence rate in the $6$--$8\,R_\oplus$ radius range. $\Delta\mathrm{ELPD}$ values are reported relative to the best model {in each row}, with standard errors. The significance in the parenthesis is $|\Delta\mathrm{ELPD}|/\mathrm{SE_\Delta}$. {Larger values indicate a stronger preference for the best model; none of the entries reaches the $2\sigma$ level except the single power law for the combined FGK sample.}}
\label{tab:loo}
\begin{tabular}{lccc}
\hline
Sample & Power-law  & Broken power-law & Power-law plus log-normal \\
\hline
FGK & $11.33\pm4.86$ {($2.3\sigma$)} & $1.84\pm1.28$ {($1.4\sigma$)} & 0 \\
F   & $1.50\pm1.56$ {($1.0\sigma$)} & $0.27\pm0.80$ {($0.3\sigma$)} & 0 \\
G   & $1.28\pm1.82$ {($0.7\sigma$)} & 0 & $0.13\pm0.92$ {($0.1\sigma$)}\\
K   & $6.86\pm4.18$ {($1.6\sigma$)} & $2.01\pm1.47$ {($1.4\sigma$)} & 0 \\
\hline
\end{tabular}
\end{table*}

\subsection{{Occurrence Rates Inside the Neptunian Desert}}\label{subsec:desert-occ}

The boundary models of Sections~\ref{subsec:desert-boundary} and \ref{subsec:ridge-savanna} describe where the desert lies, but not how empty it is in absolute terms. We now integrate the fitted occurrence-rate densities over the desert interior to obtain the number of planets per star that the desert actually contains. This is the quantity of most direct use for target selection and for comparison with population synthesis models, and to our knowledge it has not previously been measured as a function of host-star spectral type.

Because the three regions of the desert are described by three different models, we integrate each separately, over three radius ranges that together tile the full $2$--$20\,R_\oplus$ interval of this work:
\begin{enumerate}
    \item the {upper desert}, {$8 \le R_\mathrm{p} \le 20\,R_\oplus$}, obtained by integrating the suppressed density $\Gamma = 10^{\lambda(1-A)}$ of Equation~\ref{eq:upper-boundary} over the region interior to the fitted two-segment boundary{ (see below regarding the radius range)};
    \item the {lower desert}, $2 \le R_\mathrm{p} \le 6\,R_\oplus$, obtained by integrating the 2D power-law model of Equation~\ref{eq:2d-powerlaw} over the region interior to the $\Gamma = 0.1$ contour that defines the lower boundary{;}
    \item the {middle desert}, $6 \le R_\mathrm{p} \le 8\,R_\oplus$, obtained from the broken power-law fit of Section~\ref{subsec:1d-method} integrated over periods shorter than the fitted break period $P_\mathrm{b}$, which we take to delimit the desert within this strip{.}
\end{enumerate}
In each case the integral is evaluated for every posterior sample, so the quoted uncertainties propagate the full boundary and amplitude uncertainties. Table~\ref{tab:desert-occ} collects the results. Because the three models are fitted independently to disjoint radius ranges, and therefore to disjoint sets of planet candidates, their posteriors carry no statistical cross-correlation, and the three components may simply be added and their uncertainties combined in quadrature to obtain the total. The only correlation between them enters through the shared stellar sample and completeness model, which shifts all three in the same direction and is not captured by either the individual or the combined uncertainties.

We draw attention to one deliberate mismatch between the fitting and integration ranges. The two-segment upper boundary is fitted over $8 \le R_\mathrm{p} \le 16\,R_\oplus$ (Section~\ref{subsubsec:upper-boundary}), the range over which the boundary is actually resolved, but we integrate the resulting occurrence-rate density over $8 \le R_\mathrm{p} \le 20\,R_\oplus$ so that the three regions tile the full radius range of this paper and the total is directly comparable with the overall occurrence rates of Section~\ref{subsubsec:2d-occ}. This does not involve extrapolating a fitted slope: the boundary is vertical at constant period $P_0$ for $R_\mathrm{p} > R_b$, and the fitted break radius satisfies $R_b < 16\,R_\oplus$ in all four samples (Table~\ref{tab:upper-boundary}), so the boundary above $16\,R_\oplus$ is simply the continuation of that vertical segment. It does, however, assume that the desert floor density $10^{\lambda(1-A)}$ inferred over $8$--$16\,R_\oplus$ continues to apply at larger radii, where our sample contains very few planets. 

Three features stand out. First, the desert is genuinely empty. Summed over $2$--$20\,R_\oplus$, it contains $0.069^{+0.045}_{-0.011}$, $0.093^{+0.046}_{-0.022}$, and $0.150^{+0.033}_{-0.026}$ planets per 100 F-, G-, and K-type stars, which is $0.6$, $0.9$, and $1.1$ per cent of the total close-in occurrence rate measured in Section~\ref{subsubsec:2d-occ} for the same hosts. Only about one close-in planet in a hundred lies inside the desert.

Second, the upper desert shows a clear spectral-type dependence in the same sense as, and with greater contrast than, the boundary parameters. The K-type upper desert contains $0.0520^{+0.0251}_{-0.0200}$ planets per 100 stars, a factor of $5.0$ more than the F-type value ($2.0\sigma$) and $6.1$ more than the G-type value ($1.9\sigma$), while the F- and G-type values are indistinguishable from one another. The middle desert behaves the same way, with the K-type rate a factor of $4.9$ above the F-type rate, although here the F-type uncertainty is large enough that the difference is only $0.95\sigma$. This is not simply a reflection of the higher overall occurrence rate around K-type stars: normalising by the hot-Jupiter rate of each subsample (Section~\ref{subsec:hj-occ}, below), the upper desert holds $3.5$, $1.7$, and $7.8$ per cent as many planets as the hot Jupiter population does for F-, G-, and K-type hosts. Relative to the giant planets that sit just outside it, the desert is thus two to five times more populated around K-type stars.

Third, the lower desert runs the other way, decreasing monotonically from $0.0467^{+0.0014}_{-0.0013}$ (F) through $0.0464^{+0.0009}_{-0.0010}$ (G) to $0.0404^{+0.0016}_{-0.0016}$ (K) planets per 100 stars, although the total range is only $16$ per cent. This apparent reversal requires care and does not, in our view, indicate that the low-radius desert is emptier around cooler stars. Two considerations bear on it. The integration region is defined by each subsample's own fitted boundary, and that boundary lies at larger radii and shorter period for K-type hosts (Section~\ref{subsec:desert-boundary}), so the K-type region is intrinsically smaller in area; a smaller integral therefore does not imply a lower density. 
In addition, the quoted uncertainties propagate only the parameter uncertainties of a smooth power law whose indices are dominated by the much more numerous planets outside the desert, so they are far smaller than the true systematic uncertainty and the formal $3.1\sigma$ significance of the F--K difference should not be taken at face value. Both effects act in the same direction, and we therefore regard the lower-desert integrals as measurements of the shrinking extent of the desert towards later spectral types rather than of its emptiness.

Taken together, the three regions give a coherent picture: the total desert occurrence rate rises monotonically by a factor of $2.2$ from F- to K-type hosts, significant at $1.6\sigma$, and this rise is driven entirely by the upper and middle desert, i.e.\ by intermediate-radius and giant planets at $R_\mathrm{p} \gtrsim 6\,R_\oplus$, while the low-radius desert shrinks slightly in extent. We discuss the implications in Section~\ref{subsec:discussion-desert}.

Because our desert regions are delimited by our own fitted boundaries, it is important to establish how far the resulting rates depend on that choice. Paper~I provides a clean control, because it analyses the identical magnitude-limited stellar sample and the identical candidate list, which the present work merely partitions by spectral type. Integrating those occurrence rates over the fixed desert region defined by \citet{castro-gonzalezMappingExoNeptunianLandscape2024}, rather than over boundaries fitted to the data, gives an FGK desert occurrence rate of $0.08 \pm 0.01$ planets per 100 stars \citep[][]{cuiDemographicsCloseinTESS2026}. Since the stars, the candidates, the completeness model, and the occurrence-rate framework are all common to the two measurements, the comparison isolates the effect of the desert definition alone. The Paper~I value is $16$ per cent lower than our self-consistent FGK value of $0.095^{+0.010}_{-0.009}$, a difference of $1.1\sigma$. The two therefore agree within their uncertainties, and we conclude that the total desert occupancy is robust at the ${\sim}20$ per cent level against the definition of the desert, well below the factor-of-$2.2$ spectral-type variation we are attempting to measure. We note that only the self-consistent definition can be applied per spectral type, since a fixed literature region would by construction remove the spectral-type dependence of the boundaries themselves.

As a further consistency check, the star-weighted mean of our F-, G-, and K-type totals, using the subsample sizes of Section~\ref{subsec:sample}, is $0.091$ planets per 100 stars, within $5$ per cent of the $0.095$ obtained by fitting the combined FGK sample directly. The spectral-type decomposition therefore reproduces the combined result, {although this is not necessarily expected}, since the four sets of boundaries are fitted independently.

\begin{table*}
\centering
\caption{{Occurrence rates inside the Neptunian desert, in planets per 100 stars, obtained by integrating the fitted occurrence-rate densities over the desert interior. The three radius ranges tile $2$--$20\,R_\oplus$ and are integrated using the upper-boundary model (Section~\ref{subsubsec:upper-boundary}), the 1D broken power law (Section~\ref{subsec:1d-method}), and the 2D power-law lower-boundary model (Section~\ref{subsubsec:lower-boundary}), respectively. The upper-boundary model is fitted over $8$--$16\,R_\oplus$ and integrated over $8$--$20\,R_\oplus$ along its vertical segment; see Section~\ref{subsec:desert-occ}. $P_\mathrm{b}$ is the break period of the 1D broken power law, which delimits the middle desert. The total is the sum of the three components, with uncertainties combined in quadrature; this is valid because the three models are fitted to disjoint radius ranges and hence to disjoint sets of planet candidates.}}
\label{tab:desert-occ}
\renewcommand{\arraystretch}{1.3}
\begin{tabular}{lccccc}
\hline
{Sample} & {Upper desert} & {$P_\mathrm{b}$ (d)} & {Middle desert} & {Lower desert} & {Total} \\
      & {$8$--$20\,R_\oplus$} &  & {$6$--$8\,R_\oplus$} & {$2$--$6\,R_\oplus$} & {$2$--$20\,R_\oplus$} \\
\hline
{F}   & {$0.0105^{+0.0053}_{-0.0047}$} & {$4.38^{+8.74}_{-2.58}$} & {$0.0118^{+0.0445}_{-0.0104}$} & {$0.0467^{+0.0014}_{-0.0013}$} & {$0.069^{+0.045}_{-0.011}$} \\
{G}   & {$0.0086^{+0.0116}_{-0.0045}$} & {$6.48^{+5.49}_{-2.39}$} & {$0.0384^{+0.0448}_{-0.0217}$} & {$0.0464^{+0.0009}_{-0.0010}$} & {$0.093^{+0.046}_{-0.022}$} \\
{K}   & {$0.0520^{+0.0251}_{-0.0200}$} & {$4.24^{+0.72}_{-0.48}$} & {$0.0573^{+0.0221}_{-0.0173}$} & {$0.0404^{+0.0016}_{-0.0016}$} & {$0.150^{+0.033}_{-0.026}$} \\
{FGK} & {$0.0305^{+0.0072}_{-0.0069}$} & {$4.27^{+0.69}_{-0.61}$} & {$0.0190^{+0.0066}_{-0.0056}$} & {$0.0457^{+0.0007}_{-0.0008}$} & {$0.095^{+0.010}_{-0.009}$} \\
\hline
{FGK$^{\dagger}$} & {---} & {---} & {---} & {---} & {$0.08\pm0.01$} \\
\hline
\end{tabular}

{\raggedright $^{\dagger}$ Value from Paper~I \citep[][]{cuiDemographicsCloseinTESS2026}, obtained by integrating the same \textit{TESS}-SPOC occurrence rates over the fixed desert region defined by \citet{castro-gonzalezMappingExoNeptunianLandscape2024} rather than over the boundaries fitted in this work. The agreement with our FGK value provides a check that the total desert occupancy is not strongly sensitive to how the desert is delimited.}
\end{table*}

\subsection{Hot Jupiter Occurrence Rate by Spectral Type}\label{subsec:hj-occ}
\begin{figure}
    \centering
    \includegraphics[width=\linewidth]{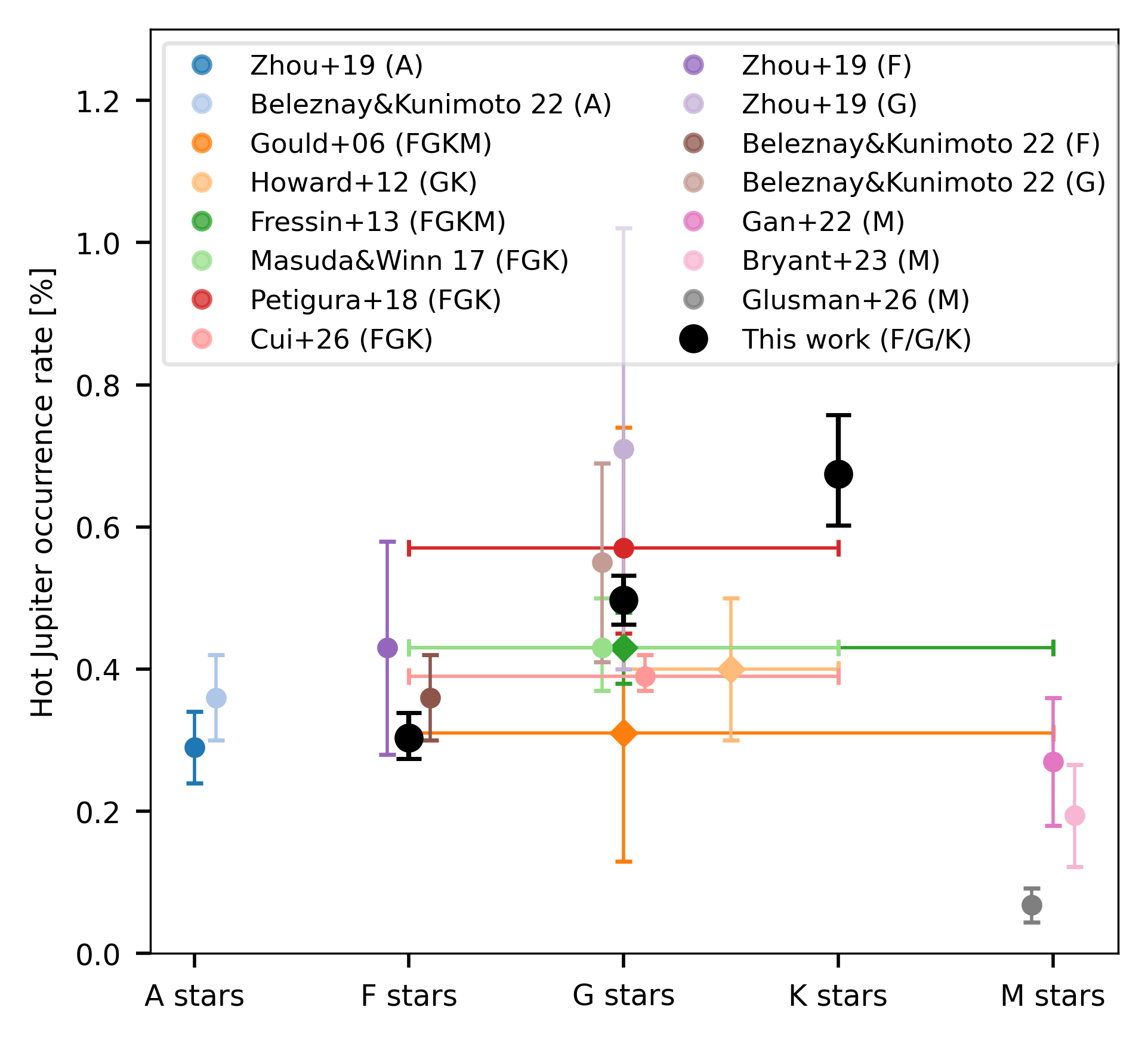}
    \caption{Occurrence rates of transiting hot Jupiters for different stellar types. Literature results are shown in different colours, and our results are shown as black dots. For results spanning more than one spectral type, we use large error bars to indicate the corresponding spectral-type range. {Small random jitters are applied along the x-axis to the literature results for clarity.}}
    \label{fig:hj-occ}
\end{figure}

Throughout this section we define hot Jupiters as planets with $8 < R_\mathrm{p} < 20\,R_\oplus$ and $0.8 < P < 10$\,d, following Paper~I, and we integrate our occurrence-rate posteriors over that region. Published occurrence rates of transiting hot Jupiters used for comparison are summarised in Table~\ref{tab:hot-jupiters-occ}. Measurements for FGK stars generally lie between approximately $0.3$ and $0.6$ per cent. We measure occurrence rates of {$0.30^{+0.04}_{-0.03}$} per cent for F-type stars, $0.50\pm0.03$ per cent for G-type stars, and {$0.67^{+0.08}_{-0.07}$} per cent for K-type stars. To our knowledge, this is the first occurrence-rate estimate specifically for hot Jupiters around K-type stars. As shown in Fig.~\ref{fig:hj-occ}, the occurrence rate in our sample increases from F- to K-type stars and reaches its maximum in the K-type bin{; unlike the integrated rates of Section~\ref{subsubsec:2d-occ}, this trend is monotonic and statistically significant, the F--G, G--K, and F--K differences corresponding to $4.0\sigma$, $2.2\sigma$, and ${>}4\sigma$, respectively}. We restrict this comparison to transiting planets because occurrence rates inferred from transit and radial-velocity surveys are subject to different selection effects and have been reported to differ systematically {\citep[see][for a recent review]{winnPlanetOccurrenceDoppler2024}}.

Previous spectral-type-resolved \textit{TESS} studies found an increasing hot-Jupiter occurrence rate from A- to G-type stars, whereas the lower rates measured for M dwarfs suggest that the occurrence rate may peak near solar-mass stars \citep{zhouTwoNewHATNet2019,beleznayExploringDependenceHot2022,ganOccurrenceRateHot2023,bryantOccurrenceRateGiant2023}. {Combining the M-dwarf measurements of \citet{ganOccurrenceRateHot2023} and \citet{bryantOccurrenceRateGiant2023} with our K-type value implies that the peak, if real, must lie within the K spectral class rather than at solar mass.} Although our measurements instead peak in the K-type bin, this result may partly reflect the stellar population within that bin. The \textit{TESS}-SPOC target sample is magnitude limited, and its K-type population is dominated by relatively hot stars with effective temperatures near {$5000\,\mathrm{K}$}, as illustrated in Fig.~\ref{fig:stellar-temperature}. By comparison, our volume-limited 100\,pc sample, which contains a larger proportion of cooler K-type stars, yields a hot-Jupiter occurrence rate of $0.42^{+0.16}_{-0.12}$ per cent \citep{cuiDemographicsCloseinTESS2026}. These results demonstrate that comparisons between broad spectral-type bins should be treated cautiously, because differences in the underlying distributions of effective temperature, stellar mass, and other stellar properties can substantially affect the inferred occurrence rates. A more informative approach would be to model hot-Jupiter occurrence continuously as a function of effective temperature or stellar mass, although such an analysis would require a larger sample and additional assumptions regarding the functional form of the dependence.

Two further considerations bear on the interpretation of this trend, both of which act in the same direction and would, if corrected for, reduce the F-to-K contrast. First, hot-Jupiter occurrence is a steeply increasing function of host-star metallicity \citep[e.g.][]{guoMetallicityDistributionHot2017, osbornInvestigatingPlanetMetallicity2020, rampalliDisentanglingMetallicityEffects2025}, and because our sample is magnitude limited, the F-type subsample is drawn from a substantially larger volume than the K-type subsample and therefore samples a different, on average more metal-poor, mixture of Galactic populations. Second, close stellar companions suppress planet occurrence, and the close-binary fraction increases with primary mass \citep[][]{moeMindYourPs2017, moeImpactBinaryStars2021}; our RUWE modelling (Section~\ref{subsec:completeness}) accounts for the effect of companions on our {detection and validation} efficiency but not for this astrophysical suppression, which would preferentially depress the F-type rate.

\begin{table*}
\centering
\caption{Published occurrence rates of transiting hot Jupiters and short-period giant planets used for comparison{, together with the measurements from this work (bottom block)}.}
\label{tab:hot-jupiters-occ}
\begin{tabular}{llcc}
\hline
Reference & Survey & Spectral type & Occurrence rate (per cent)\\
\hline
\citet{gouldFrequencyHotJupiters2006}                    & OGLE-III   & FGKM & $0.31^{+0.43}_{-0.18}$ \\
\citet{howardPLANETOCCURRENCE0252012}                   & \textit{Kepler} & GK   & $0.40\pm0.10$ \\
\citet{fressinFALSEPOSITIVERATE2013}                  & \textit{Kepler} & FGKM & $0.43\pm0.05$ \\
\citet{masudaReassessmentNullResult2017}               & \textit{Kepler} & FGK  & $0.43^{+0.07}_{-0.06}$\\
\citet{petiguraCaliforniaKeplerSurveyIV2018}                 & CKS/\textit{Kepler} & FGK & $0.57^{+0.14}_{-0.12}$ \\
\citet{cuiDemographicsCloseinTESS2026}                      & \textit{TESS} & FGK  & $0.39^{+0.03}_{-0.02}$ \\
\citet[][100\,pc volume-limited]{cuiDemographicsCloseinTESS2026}                      & \textit{TESS} & FGK  & $0.42^{+0.16}_{-0.12}$ \\
\citet{zhouTwoNewHATNet2019}                     & \textit{TESS} & A & $0.26\pm0.11$ \\
\citet{zhouTwoNewHATNet2019}                     & \textit{TESS} & F & $0.43\pm0.15$ \\
\citet{zhouTwoNewHATNet2019}                     & \textit{TESS} & G & $0.71\pm0.31$ \\
\citet{beleznayExploringDependenceHot2022}         & \textit{TESS} & A & $0.29\pm0.05$ \\
\citet{beleznayExploringDependenceHot2022}         & \textit{TESS} & F & $0.36\pm0.06$ \\
\citet{beleznayExploringDependenceHot2022}         & \textit{TESS} & G & $0.55\pm0.14$ \\
\citet{ganOccurrenceRateHot2023}                      & \textit{TESS} & M & $0.27\pm0.09$ \\
\citet{bryantOccurrenceRateGiant2023}                   & \textit{TESS} & M & $0.194\pm0.072$ \\
\citet[][100\,pc volume-limited]{glusmanSearchingGEMSOccurrence2026}                   & \textit{TESS} & M & $0.068\pm0.024$ \\
\hline
{This work}                     & {\textit{TESS}} & {F} & {$0.30^{+0.04}_{-0.03}$} \\
{This work}                     & {\textit{TESS}} & {G} & {$0.50\pm0.03$} \\
{This work}                     & {\textit{TESS}} & {K} & {$0.67^{+0.08}_{-0.07}$} \\
\hline
\end{tabular}
\end{table*}

\section{Discussion}\label{sec:discussion}

\subsection{Comparison with \textit{Kepler} Occurrence Rates}\label{subsec:kepler-comparison}
Here, we present a direct comparison with \textit{Kepler} using the occurrence-rate results from \citet{kunimotoSearchingEntiretyKepler2020}, which provide maps in the period--radius plane separately for F-, G-, and K-type stars {and which adopt the same effective-temperature boundaries as we do, making the comparison as close to like-for-like as the two surveys allow}. Because the \textit{Kepler} results cover a larger parameter space, we restrict the comparison to the overlapping region ($0.78 < P < 12.5$\,d, $2 < R_\mathrm{p} < 16\,R_\oplus$). To enable a detailed bin-by-bin comparison, we fit Gamma distributions {to their tabulated results}, using the median values and the upper (84.1st percentile) and lower (15.9th percentile) uncertainties listed in Table~A1 of \citet{kunimotoSearchingEntiretyKepler2020}. For bins with only upper limits, we fit the distributions by fixing $\alpha = 1$, such that the Gamma distribution reduces to an exponential distribution. {The Gamma family is a natural choice here because it is the conjugate form for a Poisson rate and is guaranteed to be positive, but we caution that reconstructing a posterior from three quantiles is necessarily approximate, and the reconstruction is least reliable precisely in the sparsely populated bins where the two surveys differ most.}

\begin{figure}
    \centering
    \includegraphics[width=\linewidth]{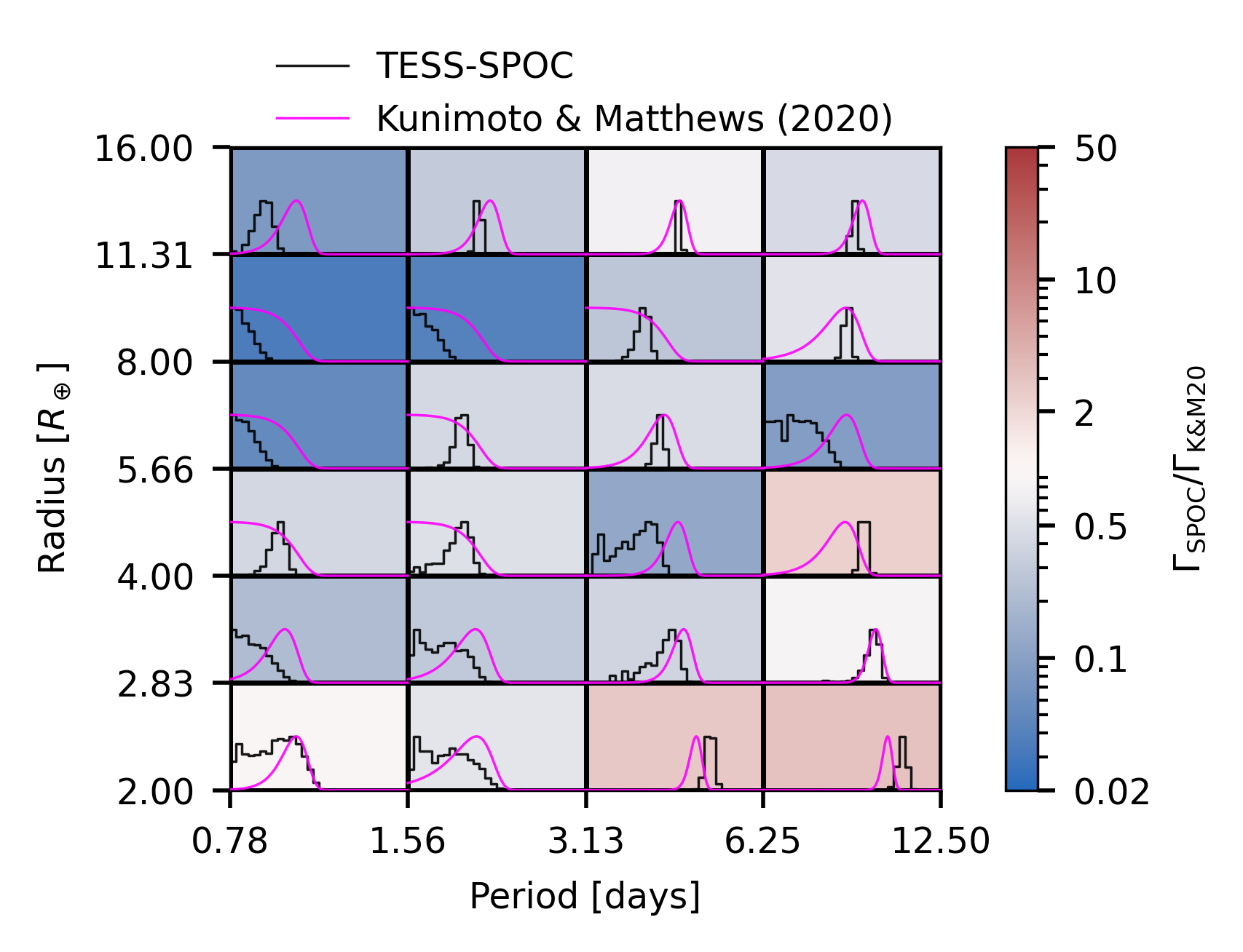}
    \caption{Posterior distributions of planetary occurrence rates of F-type stars for both {\textit{TESS}}-SPOC {(black)} and \citet{kunimotoSearchingEntiretyKepler2020} {(fuchsia)}. In each cell, the {$x$-axis spans} occurrence rates {from} $1\times10^{-4}$ per cent to 50 per cent on a logarithmic scale. The heights of the distributions are {scaled equally} in each cell. The background diverging colour map indicates the ratio of median occurrence rates between the two samples. The colour scale is manually centred {on unity}; the actual minimum ratio is approximately 0.03.}
    \label{fig:f-type-kepler-comparison}
\end{figure}

\begin{figure}
    \centering
    \includegraphics[width=\linewidth]{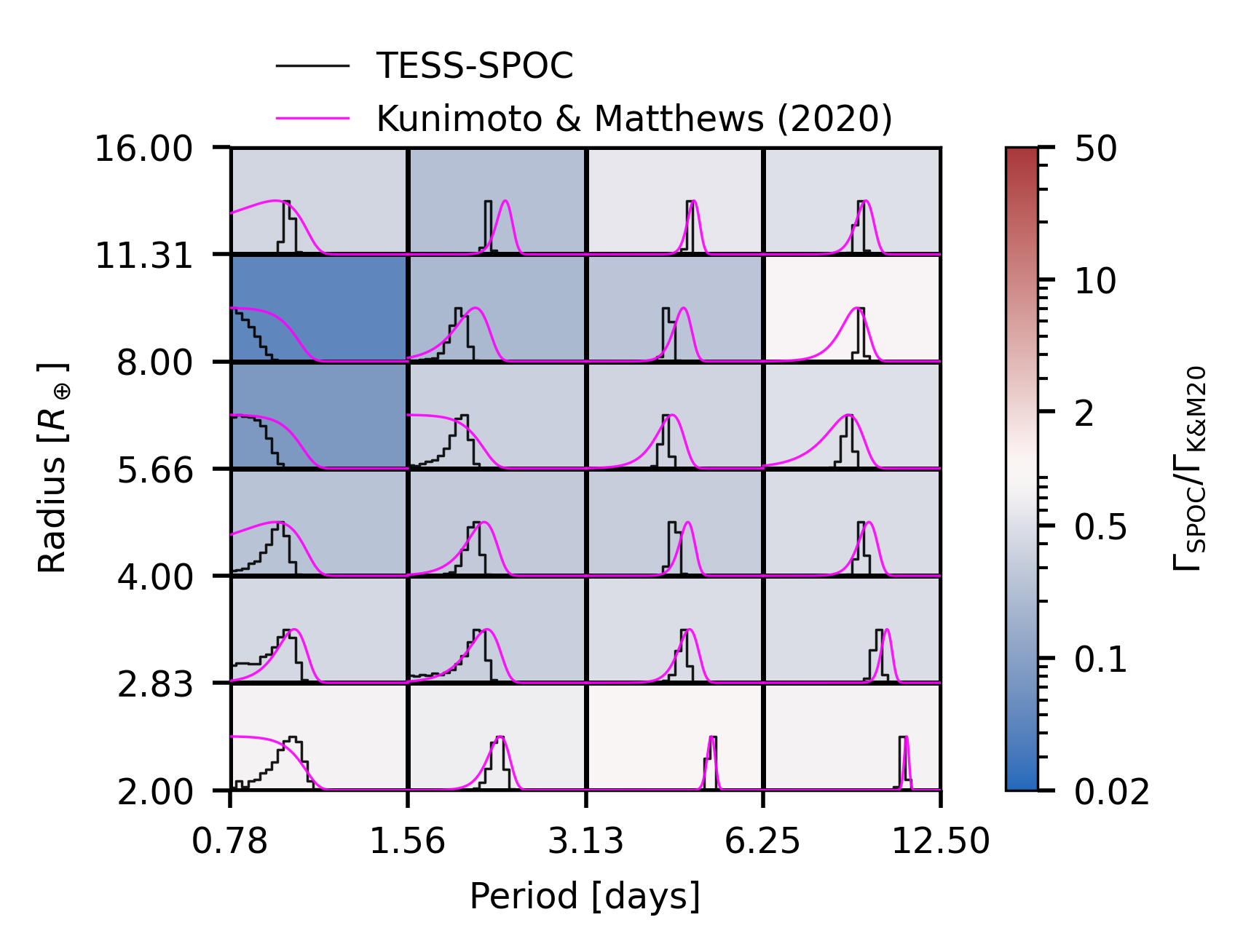}
    \caption{{Same as Fig.~\ref{fig:f-type-kepler-comparison}}, but for G-type stars. The actual minimum ratio is approximately 0.05.}
    \label{fig:g-type-kepler-comparison}
\end{figure}

\begin{figure}
    \centering
    \includegraphics[width=\linewidth]{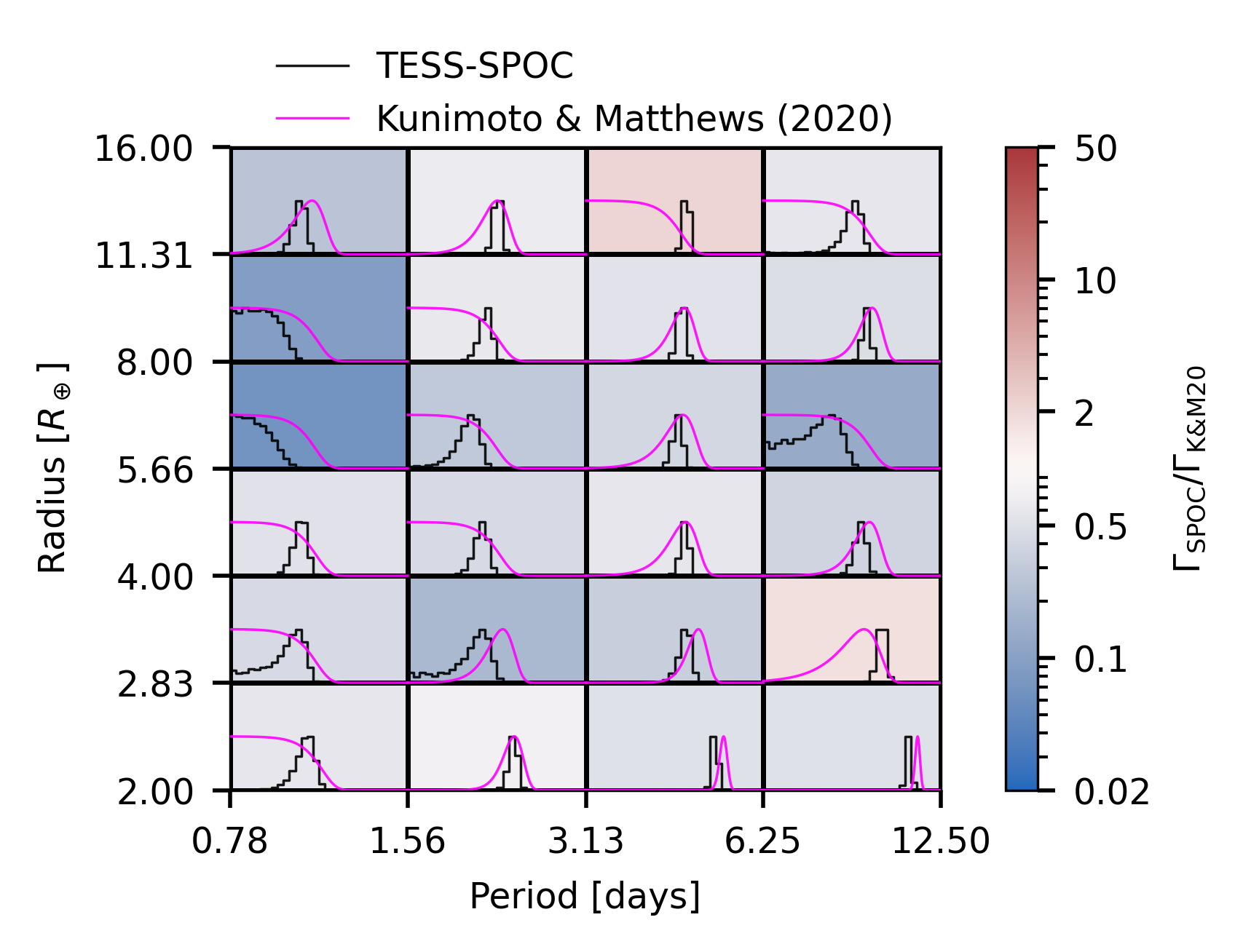}
    \caption{{Same as Fig.~\ref{fig:f-type-kepler-comparison}}, but for K-type stars. The actual minimum ratio is approximately 0.07.}
    \label{fig:k-type-kepler-comparison}
\end{figure}

Given the same grid configuration as \citet{kunimotoSearchingEntiretyKepler2020}, we run our Bayesian model as introduced in {Section~\ref{subsec:occ-method}}. {Figs.~\ref{fig:f-type-kepler-comparison}}, \ref{fig:g-type-kepler-comparison}, and \ref{fig:k-type-kepler-comparison} show our {\textit{TESS}}-SPOC occurrence-rate posterior distributions (black lines) in each bin and the fitted {\textit{Kepler}} occurrence-rate distributions (fuchsia lines). The colour scale represents the ratio of the {\textit{TESS}}-SPOC occurrence rate to the {\textit{Kepler}} occurrence rate. Accordingly, bluish and reddish colours indicate {\textit{TESS}}-SPOC occurrence rates that are, respectively, lower and higher than those from {\textit{Kepler}}.

These posterior distributions show that our occurrence rates are generally consistent with the {\textit{Kepler}} results within uncertainties, but are significantly tighter in most bins than those of \citet{kunimotoSearchingEntiretyKepler2020}, especially in the well-constrained bins. In the desert region, our posterior places even lower upper limits, mainly due to the larger stellar sample size of {\textit{TESS}}-SPOC.

{The agreement is not trivial, and it is worth being explicit about why. The two measurements share almost nothing beyond the definition of the spectral-type bins: they use different telescopes, photometric precisions, cadences, and time baselines; different transit-search and vetting algorithms; different stellar catalogues, and hence different planet radius scales; and different statistical frameworks, \citet{kunimotoSearchingEntiretyKepler2020} using approximate Bayesian computation \citep[][]{kunimotoComparingApproximateBayesian2020} against our hierarchical Poisson-process model. The fact that the two nonetheless agree bin by bin over more than two decades in occurrence-rate density suggests that neither is dominated by an unmodelled completeness systematic, which is the principal worry for occurrence-rate measurements at short periods.

Where the two do differ, the differences are informative rather than random. Our rates fall below the \textit{Kepler} values by up to a factor of ${\sim}30$ in the most extreme bins, and these bins are concentrated at intermediate-to-large radii ($5.7 \lesssim R_\mathrm{p} \lesssim 11\,R_\oplus$) and at the shortest periods, that is, inside the Neptunian desert. In these bins, \textit{Kepler} detected no objects. Our posteriors are themselves one-sided upper limits (visible as the monotonically declining black curves in Figs.~\ref{fig:f-type-kepler-comparison}--\ref{fig:k-type-kepler-comparison}), and the \textit{Kepler} posterior is prior dominated, so the ratio of medians is not a meaningful test of consistency; what matters is that our upper limits lie below theirs, which is the expected consequence of a stellar sample roughly an order of magnitude larger. The bins where our rates exceed the \textit{Kepler} values are located in the lower-right corner of our parameter space as they are sensitive to the relatively low completeness there.

Finally, we emphasise a difference in what the two samples represent. The \textit{Kepler} field samples a single, relatively distant line of sight towards the Galactic plane, whereas our sample is all-sky and, being magnitude limited, is drawn from a much more local volume. Any dependence of planet occurrence on Galactic environment, metallicity, or age therefore enters the two measurements differently \citep[e.g.][]{rampalliDisentanglingMetallicityEffects2025, sayeedExoplanetOccurrenceRate2025}. That we find agreement at the current precision places a useful, if indirect, limit on how strong such environmental dependences can be for close-in planets.}

\subsection{Stellar Type Dependence of Desert Boundaries and Occurrence}\label{subsec:discussion-desert}

\begin{figure}
    \centering
    \includegraphics[width=\linewidth]{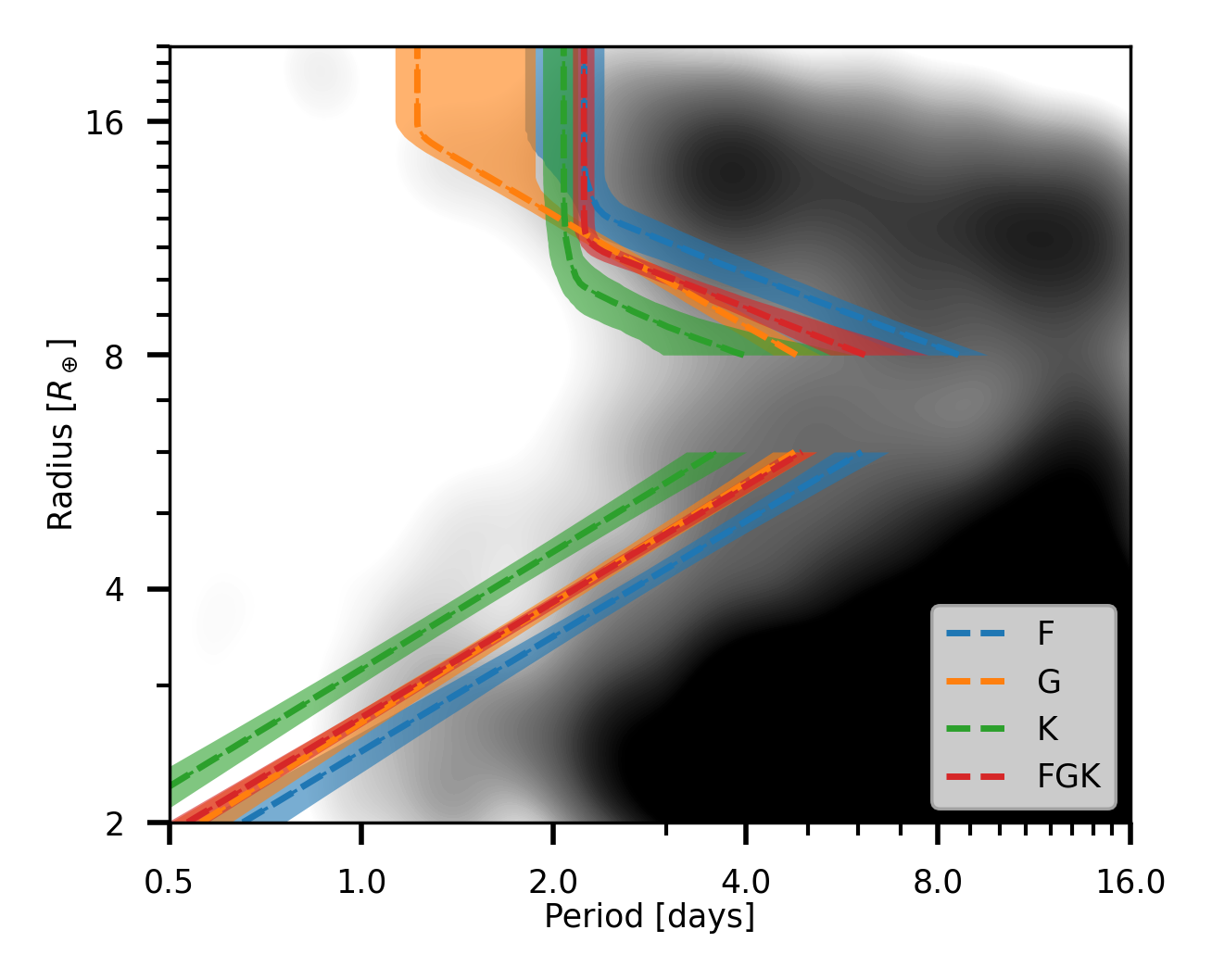}
    \caption{Fitted Neptunian desert boundaries for the F- (blue), G- (orange), and K-type (green) subsamples, together with the combined FGK sample (red), overlaid in the period--radius plane. Dashed lines show the two-segment upper boundary (Section~\ref{subsubsec:upper-boundary}) and $\Gamma = 0.1$ lower-boundary contour (Section~\ref{subsubsec:lower-boundary}); shaded bands indicate the 16th--84th percentile intervals of posteriors. The wKDE of the overall FGK sample is shown as the background.}
    \label{fig:boundary-comparison}
\end{figure}

Fig.~\ref{fig:boundary-comparison} collects the fitted upper and lower boundaries of all four samples in a single period--radius plane. The upper boundary does not translate rigidly with spectral type: it pivots about its inner edge and shortens. The near-vertical short-period segment shows no ordered dependence on spectral type, $P_0 = 2.24^{+0.17}_{-0.39}$\,d (F), $1.22^{+0.67}_{-0.09}$\,d (G), and $2.08^{+0.11}_{-0.16}$\,d (K): the F and K values, which span the widest lever arm in $T_\mathrm{eff}$, agree to within a factor of $1.08$ ($0.4\sigma$), and the lower G-type value is not ordered between them, so it cannot be read as a trend. The G-type posterior is strongly asymmetric and is the least secure of the three (Section~\ref{subsec:desert-boundary}); all three values are mutually consistent within ${\sim}1.3\sigma$. By contrast the outer terminus retreats monotonically, from $P_1 = 8.62^{+0.95}_{-1.57}$\,d (F) through $4.81^{+0.42}_{-0.39}$\,d (G) to $4.05^{+2.00}_{-1.04}$\,d (K), a factor of $2.1$. The lower boundary meanwhile lifts: both fitted indices shift in the same direction from F to K (Table~\ref{tab:lower-boundary}), moving the $\Gamma = 0.1$ contour upward in radius and inward in period, so that sub-Neptunes around cooler hosts retain their envelopes at smaller separations and to larger sizes. The desert therefore closes from both above and below towards later spectral types.

As discussed in Section {\ref{subsec:desert-occ}}, planet occurrence inside the desert complements the stellar type dependence of its boundaries: K stars host ${\sim}2{-}5{\times}$ more planets inside the desert, relative to giant planets just exterior to the desert, than F {and G} stars. This increase in occurrence inside the desert toward lower mass stars is primarily concentrated in the middle/upper desert, i.e. ultrahot gas-rich planets. Below we outline how these desert boundary and occurrence measurements may inform theoretical interpretation of the migratory and mass loss processes sculpting the desert.


That the two segments of the upper boundary behave differently is the clearest indication in our data that they are sculpted by differing processes. The invariance of $P_0$ is suggestive of a tidally-set edge: the tidal disruption orbital period depends only on the mean density of the planet, the stellar mass cancelling exactly \citep[e.g.][]{castro-gonzalezNeptunianRidgeNatural2026}, so a boundary established by tidal disruption during high-eccentricity migration \citep[][]{matsakosORIGINSUBJOVIANDESERT2016,owenPhotoevaporationHigheccentricityMigration2018} or potentially Roche-lobe overflow (RLO) should be largely independent of spectral type \cite[with the caveat that different stellar effective temperatures may alter RLO mass loss rates via the gas sound speed/mean molecular weight;][]{jacksonNewModelRoche2017,hallattSheddingLightDesert2026}. On the other hand, the fact that the outer locus of the desert ($P_1$) swells outward in period toward larger stellar masses runs counter to this expectation from tidal disruption. One potential explanation for this discrepancy with the tidal survival hypothesis is that the upper boundaries are post-processed by stellar irradiation-driven mass loss, after planets are initially emplaced at the (stellar mass-independent) tidal boundary. Future work is needed to quantitatively assess whether such mass loss (photoevaporative or bolometrically-catalyzed, or a combination of the two) may indeed reproduce the stellar mass-dependent outer locus of the desert ($P_1$).

Irradiation-driven mass loss may also explain the lower boundary's rise toward lower mass stars. 
\citet{mcdonaldSubNeptuneDesertIts2019} argued that the sub-Neptune desert is controlled by lifetime-integrated X-ray fluence, which varies far more weakly across spectral types than instantaneous irradiation does because the saturated high-energy phase lasts progressively longer towards later types; a boundary set by it should shift inward modestly, which is what we measure. The shift in the observed desert boundaries across stellar types is examined further in \citet{osbornConfirmationHotSuperNeptune2026} and King et al. (submitted). They find the shifts to be consistent with variations in the XUV irradiation history across stellar types, and their investigation of the lower boundary in cumulative XUV space is consistent with photoevaporation sculpting that part of the desert.

Planet occurrence inside these desert boundaries further constrains how irradiation-driven mass loss and tidal survival sculpt the desert. Empirical \citep{vissapragadaHottestNeptunesOrbit2025} and theoretical work suggests that the desert may be populated with the remnants hot Jupiters, destroyed either via Roche lobe overflow by tidal inspiral \citep[i.e. bolometrically-driven, tidally-accelerated Parker wind mass loss;][]{hallattSheddingLightDesert2026}, or tidal shocks during high eccentricity migration \cite[][]{zanazziDynamicalTidesHighEccentricity2026}. Our measurements support this ``top-down" hypothesis for the origin of planets in the desert in two ways. First, occurrence of planets with radii $> 6\,R_\oplus$ in the desert relative to hot Jupiter occurrence of the same subsample is just $3.5$, $1.7$, and $7.8$ per cent for F, G, and K hosts, as qualitatively expected if desert planets represent the minority of hot Jupiters that undergo extreme mass loss. Second, the desert's inner edge is sharp ($\delta = 0.02$--$0.06$\,dex), as expected of a destruction threshold rather than of a smoothly declining distribution. Our finding that K stars exhibit enhanced occurrence of ultrahot gas-rich planets inside the desert relative to hot Jupiters just beyond could further point toward more efficient tidal transport of hot Jupiters to shorter periods via stellar tides \citep[e.g.][]{barkerTidalDissipationEvolving2020}. Future work with an expanded planet sample may also measure planet occurrence in the desert above/below the Kraft break, to further differentiate the RLO vs. high eccentricity migration mechanisms \cite[][]{hallattRevealingOriginDesert2026}.

Alternatively, if the desert is partly a gap in the population as delivered rather than an erosion feature, the relevant scalings are those of disc mass, disc lifetime, and the timing of runaway accretion, all of which vary with stellar mass and could shift the boundary in period with no counterpart in irradiation space \citep[e.g.][]{kennedyPlanetFormationStars2008, ribasProtoplanetaryDiskLifetimes2015}. And because the upper boundary is also the lower envelope of the hot-Jupiter population, whose occurrence rises steeply with metallicity, the more metal-poor mixture sampled by our magnitude-limited F-type subsample (Section~\ref{subsec:hj-occ}) could contribute to the apparent motion of $P_1$ without any change in the sculpting physics. Settling this will require measurements of a different kind, since these mechanisms predict similar boundary locations but divergent properties for the planets inside the desert: obliquities clustered near alignment for RLO versus broadly distributed for high-eccentricity migration \citep[][]{hallattRevealingOriginDesert2026}, together with residual eccentricities \citep[][]{bourrierDREAMOrbitalArchitecture2023}, bulk densities \citep[][]{armstrongRemnantPlanetaryCore2020,hallattSheddingLightDesert2026}, and host-star rotation and kinematics. Our contribution is to fix the boundary locations, and their spectral-type dependence, in completeness-corrected occurrence-rate space, which is what those population-level tests must be compared against.

\subsection{Future Prospects}\label{subsec:future}

Several avenues for extending this work remain. First, the addition of \textit{TESS} sectors beyond Sector~55 will increase the planet yield and allow more statistically significant detections of features such as the ridge; since the LOO differences in Table~\ref{tab:loo} scale roughly linearly with sample size, distinguishing a log-normal ridge from a broken power law at the $3\sigma$ level would require of order four times our present candidate count in the $6$--$8\,R_\oplus$ strip. Second, extending the radius range below $2\,R_\oplus$ will enable spectral-type-resolved studies of the radius valley and super-Earths, whose location is predicted to depend on stellar mass in a way that discriminates between photoevaporation and core-powered mass loss \citep[][]{owenEvaporationValleyKepler2017, fultonCaliforniaKeplerSurveyIII2017, vaneylenAsteroseismicViewRadius2018, petiguraCaliforniaKeplerSurveyRadius2022, hoDeepRadiusValley2023}. Third, incorporating stellar metallicity from spectroscopic surveys or from \textit{Gaia} XP spectra \citep[][]{andraeRobustDatadrivenMetallicities2023} will allow investigation of the joint dependence of occurrence rates on effective temperature and metallicity, disentangling stellar mass and composition effects. Fourth, the boundary models presented here can be compared directly with predictions from population synthesis codes to constrain the relative contributions of photoevaporation, core-powered mass loss, and tidal effects.

\section{Conclusion}\label{sec:conclusion}

In this second paper of our series on \textit{TESS}-SPOC exoplanet occurrence rates, we have extended the analysis of Paper~I to individual F-, G-, and K-type main-sequence stellar samples. Our main findings are as follows:

\begin{itemize}
    \item Applying the detection, vetting, and validation pipeline of Paper~I to 2,097,359 FGK dwarfs and recomputing the completeness independently for each spectral type, we obtain occurrence-rate maps in the period--radius plane for F-type, G-type, and K-type planet candidates. All maps, tables, and posterior distributions are released publicly.

    \item Integrated over $0.5 < P < 16$\,d and $2 < R_\mathrm{p} < 20\,R_\oplus$, the occurrence rates are $12.1^{+3.0}_{-2.3}$, $10.2^{+1.2}_{-1.0}$, and $13.1^{+1.4}_{-1.2}$ planets per 100 F-, G-, and K-type stars. These integrals are taken over slightly different sets of occupied bins and are mutually consistent at the $1\sigma$ level, so we do not claim a monotonic trend in the integrated rate. The spectral-type dependence is instead localised, and is clearest for sub-Neptunes at $P \gtrsim 3$\,d, where it agrees with previous \textit{Kepler}-based results.

    \item For transiting hot Jupiters ($8 < R_\mathrm{p} < 20\,R_\oplus$ and $0.8 < P < 10$\,d) we measure $0.30^{+0.04}_{-0.03}$, $0.50\pm0.03$, and $0.67^{+0.08}_{-0.07}$ per cent for F-, G-, and K-type hosts, the last being the first such measurement for K-type stars. The increase from F to K is monotonic and significant at ${>}4\sigma$, but the amplitude is sensitive to the effective-temperature composition of the magnitude-limited bins: our 100\,pc volume-limited FGK sample, which contains a larger fraction of cool K dwarfs, gives a lower rate than the K-type bin alone.
    
    \item The near-vertical, short-period upper edge of the Neptunian desert shows no significant dependence on spectral type, with $P_0 = 2.24^{+0.17}_{-0.39},\ 1.22^{+0.67}_{-0.09},\ \text{and}\ 2.08^{+0.11}_{-0.16}\,\mathrm{d}$ for F, G, and K stars, respectively. The inferred edge locations differ by only $0.4\sigma$ (F--K), $1.3\sigma$ (F--G), and $1.2\sigma$ (G--K), indicating consistency within the uncertainties.

    \item The outer terminus of the upper boundary does shift inward, from $8.62^{+0.95}_{-1.57}$\,d (F) through $4.81^{+0.42}_{-0.39}$\,d (G) to $4.05^{+2.00}_{-1.04}$\,d (K).

    \item The desert is systematically less empty around cooler stars. The contrast across the upper boundary decreases from $2.28^{+0.37}_{-0.26}$\,dex (F) to $1.29^{+0.23}_{-0.20}$\,dex (K), and the independently fitted lower boundary moves to larger radii from F to K, driven by a shallower radius index ($\beta = -4.15^{+0.35}_{-0.36}$ to $-3.53 \pm 0.27$). The two boundaries fitted to disjoint radius ranges agree on the direction of the trend.

    \item The desert interior contains $0.069^{+0.045}_{-0.011}$, $0.093^{+0.046}_{-0.022}$, and $0.150^{+0.033}_{-0.026}$ planets per 100 F-, G-, and K-type stars, i.e.\ only about one close-in planet in a hundred, and to our knowledge this is the first such measurement resolved by spectral type. The total rises by a factor of $2.2$ from F to K, driven entirely by $R_\mathrm{p} \gtrsim 6\,R_\oplus$, and is robust at the ${\sim}20$ per cent level against the choice of desert definition.

    \item Our one-dimensional occurrence-rate profiles in the $6$--$8\,R_\oplus$ strip do not require a distinct Neptunian ridge. A power law plus log-normal, representing the ridge, is preferred over a broken power law by only $1.4\sigma$ for the combined FGK sample, and the significance is even lower for the individual stellar-type subsamples.

    \item Our bin-by-bin posteriors agree with the \textit{Kepler} measurements of \citet{kunimotoSearchingEntiretyKepler2020} for all three spectral types while being substantially tighter, with the largest apparent discrepancies confined to bins in which one or both surveys report only upper limits.

\end{itemize}

Together with Paper~I, these results provide the most detailed spectral-type-resolved census of close-in planet demographics from \textit{TESS} to date, establishing an empirical framework for testing theoretical models of planet formation and atmospheric evolution as a function of host star properties.

\section*{Acknowledgements}
{We appreciate the referee’s helpful comments and suggestions, which have improved the clarity of this work.} This paper includes data collected by the \textit{TESS} mission. Funding for the \textit{TESS} mission is provided by the NASA Explorer Program. Resources supporting this work were provided by the NASA High-End Computing (HEC) Program through the NASA Advanced Supercomputing (NAS) Division at Ames Research Center for the production of the SPOC data products. This research was funded by the UKRI (Grants ST/X001121/1, EP/X027562/1). AI tools are used for some discussion, text editing, plot coding.
This work made use of {\tt numpy} \citep{harrisArrayProgrammingNumPy2020}, {\tt scipy} \citep{2020SciPy-NMeth}, {\tt astropy} \citep{theastropycollaborationAstropyCommunityPython2013,theastropycollaborationAstropyProjectBuilding2018}, {\tt matplotlib} \citep{hunterMatplotlib2DGraphics2007}, {\tt Jupyter} \citep{kluyverJupyterNotebooksPublishing2016}, {\tt arviz} \citep{martinArviZModularFlexible2026}, {\tt emcee} \citep[][]{foreman-mackeyEmceeMCMCHammer2013}, {\tt juliet} \citep[][]{espinozaJulietVersatileModelling2019}, and {\tt gaiamock} \citep[][]{el-badryGenerativeModelGaia2024}.

\section*{Data Availability}
The \textit{TESS}-SPOC data used in this paper can be found in the Mikulski Archive for Space Telescopes \citep{caldwellTESSLightCurves2020}. The occurrence-rate posterior distributions for each cell of the F-, G-, and K-type grids, together with the posteriors of the desert-boundaries, are available at \href{https://doi.org/10.5281/zenodo.21889834}{doi:10.5281/zenodo.21889834} and \url{https://github.com/ckm3/NDesPlot}.


\bibliographystyle{mnras}
\bibliography{Neptune}



\appendix

\section{Occurrence Rate Tables by Spectral Type}\label{appendix:tables}
Table~\ref{tab:TESS-SPOC 10x10 table F type} lists the occurrence rate values and uncertainties of Fig.~\ref{fig:occ-rate-F}. Table~\ref{tab:TESS-SPOC 10x10 table G type} lists the occurrence rate values and uncertainties of Fig.~\ref{fig:occ-rate-G}. Table~\ref{tab:TESS-SPOC 10x10 table K type} lists the occurrence rate values and uncertainties of Fig.~\ref{fig:occ-rate-K}.

\begin{table*}
\caption{Occurrence rates (number of planets per 100 stars) of our 10-by-10 bins of F-type stars (Fig.~\ref{fig:occ-rate-F}). Values are reported to two significant figures.}
\label{tab:TESS-SPOC 10x10 table F type}
\renewcommand{\arraystretch}{1.28}
\setlength{\tabcolsep}{4.1pt}
\begin{tabular}{ccc @{\ \vline\ } ccc}
\hline
Period ranges (days) & Radius ranges ($R_{\earth}$) & Occurrence rates (\%)
   & Period ranges (days) & Radius ranges ($R_{\earth}$) & Occurrence rates (\%) \\ 
\hline
0.50 -- 0.71 & 2.00 -- 2.52 & $0.012 [+0.020, -0.0091]$ &
0.50 -- 0.71 & 2.52 -- 3.17 & $0.0045 [+0.0048, -0.0030]$ \\
0.50 -- 0.71 & 3.17 -- 3.99 & $0.0025 [+0.0025, -0.0016]$ &
0.50 -- 0.71 & 3.99 -- 5.02 & $0.0031 [+0.0017, -0.0013]$ \\
0.50 -- 0.71 & 5.02 -- 6.32 & $0.00048 [+0.00075, -0.00036]$ &
0.50 -- 0.71 & 6.32 -- 7.96 & $0.00031 [+0.00052, -0.00024]$ \\
0.50 -- 0.71 & 7.96 -- 10.02 & $0.00029 [+0.00050, -0.00022]$ &
0.50 -- 0.71 & 10.02 -- 12.62 & $0.00027 [+0.00046, -0.00020]$ \\
0.50 -- 0.71 & 12.62 -- 15.89 & $0.00029 [+0.00049, -0.00022]$ &
0.50 -- 0.71 & 15.89 -- 20.00 & $0.0012 [+0.0020, -0.00089]$ \\
0.71 -- 1.00 & 2.00 -- 2.52 & $0.026 [+0.041, -0.019]$ &
0.71 -- 1.00 & 2.52 -- 3.17 & $0.0026 [+0.0043, -0.0019]$ \\
0.71 -- 1.00 & 3.17 -- 3.99 & $0.0011 [+0.0017, -0.00080]$ &
0.71 -- 1.00 & 3.99 -- 5.02 & $0.00067 [+0.0011, -0.00051]$ \\
0.71 -- 1.00 & 5.02 -- 6.32 & $0.00047 [+0.00077, -0.00035]$ &
0.71 -- 1.00 & 6.32 -- 7.96 & $0.00040 [+0.00064, -0.00030]$ \\
0.71 -- 1.00 & 7.96 -- 10.02 & $0.00030 [+0.00049, -0.00023]$ &
0.71 -- 1.00 & 10.02 -- 12.62 & $0.00060 [+0.00074, -0.00040]$ \\
0.71 -- 1.00 & 12.62 -- 15.89 & $0.00089 [+0.00085, -0.00051]$ &
0.71 -- 1.00 & 15.89 -- 20.00 & $0.0014 [+0.0023, -0.0010]$ \\
1.00 -- 1.41 & 2.00 -- 2.52 & $0.051 [+0.066, -0.035]$ &
1.00 -- 1.41 & 2.52 -- 3.17 & $0.0066 [+0.010, -0.0049]$ \\
1.00 -- 1.41 & 3.17 -- 3.99 & $0.0019 [+0.0030, -0.0014]$ &
1.00 -- 1.41 & 3.99 -- 5.02 & $0.0031 [+0.0026, -0.0017]$ \\
1.00 -- 1.41 & 5.02 -- 6.32 & $0.00077 [+0.0013, -0.00058]$ &
1.00 -- 1.41 & 6.32 -- 7.96 & $0.00055 [+0.00087, -0.00041]$ \\
1.00 -- 1.41 & 7.96 -- 10.02 & $0.00041 [+0.00066, -0.00031]$ &
1.00 -- 1.41 & 10.02 -- 12.62 & $0.00059 [+0.00082, -0.00042]$ \\
1.00 -- 1.41 & 12.62 -- 15.89 & $0.00079 [+0.00095, -0.00055]$ &
1.00 -- 1.41 & 15.89 -- 20.00 & $0.0017 [+0.0029, -0.0013]$ \\
1.41 -- 2.00 & 2.00 -- 2.52 & $0.050 [+0.085, -0.037]$ &
1.41 -- 2.00 & 2.52 -- 3.17 & $0.0090 [+0.014, -0.0067]$ \\
1.41 -- 2.00 & 3.17 -- 3.99 & $0.0034 [+0.0051, -0.0025]$ &
1.41 -- 2.00 & 3.99 -- 5.02 & $0.0066 [+0.0046, -0.0034]$ \\
1.41 -- 2.00 & 5.02 -- 6.32 & $0.0017 [+0.0021, -0.0012]$ &
1.41 -- 2.00 & 6.32 -- 7.96 & $0.00083 [+0.0013, -0.00062]$ \\
1.41 -- 2.00 & 7.96 -- 10.02 & $0.00048 [+0.00083, -0.00036]$ &
1.41 -- 2.00 & 10.02 -- 12.62 & $0.00042 [+0.00071, -0.00031]$ \\
1.41 -- 2.00 & 12.62 -- 15.89 & $0.0019 [+0.0014, -0.00094]$ &
1.41 -- 2.00 & 15.89 -- 20.00 & $0.0024 [+0.0035, -0.0018]$ \\
2.00 -- 2.83 & 2.00 -- 2.52 & $0.074 [+0.12, -0.056]$ &
2.00 -- 2.83 & 2.52 -- 3.17 & $0.0093 [+0.015, -0.0069]$ \\
2.00 -- 2.83 & 3.17 -- 3.99 & $0.0087 [+0.0094, -0.0057]$ &
2.00 -- 2.83 & 3.99 -- 5.02 & $0.0033 [+0.0042, -0.0024]$ \\
2.00 -- 2.83 & 5.02 -- 6.32 & $0.0044 [+0.0034, -0.0024]$ &
2.00 -- 2.83 & 6.32 -- 7.96 & $0.0015 [+0.0018, -0.0011]$ \\
2.00 -- 2.83 & 7.96 -- 10.02 & $0.00059 [+0.00096, -0.00044]$ &
2.00 -- 2.83 & 10.02 -- 12.62 & $0.0017 [+0.0017, -0.0011]$ \\
2.00 -- 2.83 & 12.62 -- 15.89 & $0.0065 [+0.0028, -0.0023]$ &
2.00 -- 2.83 & 15.89 -- 20.00 & $0.0093 [+0.0075, -0.0053]$ \\
2.83 -- 4.00 & 2.00 -- 2.52 & $0.41 [+0.41, -0.25]$ &
2.83 -- 4.00 & 2.52 -- 3.17 & $0.12 [+0.069, -0.054]$ \\
2.83 -- 4.00 & 3.17 -- 3.99 & $0.016 [+0.018, -0.011]$ &
2.83 -- 4.00 & 3.99 -- 5.02 & $0.0058 [+0.0067, -0.0041]$ \\
2.83 -- 4.00 & 5.02 -- 6.32 & $0.0038 [+0.0046, -0.0026]$ &
2.83 -- 4.00 & 6.32 -- 7.96 & $0.0065 [+0.0038, -0.0028]$ \\
2.83 -- 4.00 & 7.96 -- 10.02 & $0.0022 [+0.0024, -0.0015]$ &
2.83 -- 4.00 & 10.02 -- 12.62 & $0.0025 [+0.0027, -0.0017]$ \\
2.83 -- 4.00 & 12.62 -- 15.89 & $0.040 [+0.0070, -0.0060]$ &
2.83 -- 4.00 & 15.89 -- 20.00 & $0.013 [+0.016, -0.0093]$ \\
4.00 -- 5.66 & 2.00 -- 2.52 & $1.2 [+0.89, -0.62]$ &
4.00 -- 5.66 & 2.52 -- 3.17 & $0.20 [+0.13, -0.10]$ \\
4.00 -- 5.66 & 3.17 -- 3.99 & $0.036 [+0.032, -0.023]$ &
4.00 -- 5.66 & 3.99 -- 5.02 & $0.010 [+0.012, -0.0070]$ \\
4.00 -- 5.66 & 5.02 -- 6.32 & $0.0076 [+0.0083, -0.0053]$ &
4.00 -- 5.66 & 6.32 -- 7.96 & $0.017 [+0.0072, -0.0060]$ \\
4.00 -- 5.66 & 7.96 -- 10.02 & $0.0046 [+0.0035, -0.0023]$ &
4.00 -- 5.66 & 10.02 -- 12.62 & $0.0019 [+0.0025, -0.0014]$ \\
4.00 -- 5.66 & 12.62 -- 15.89 & $0.038 [+0.0077, -0.0067]$ &
4.00 -- 5.66 & 15.89 -- 20.00 & $0.0094 [+0.015, -0.0070]$ \\
5.66 -- 8.00 & 2.00 -- 2.52 & $0.73 [+1.1, -0.53]$ &
5.66 -- 8.00 & 2.52 -- 3.17 & $0.66 [+0.26, -0.22]$ \\
5.66 -- 8.00 & 3.17 -- 3.99 & $0.066 [+0.067, -0.045]$ &
5.66 -- 8.00 & 3.99 -- 5.02 & $0.028 [+0.022, -0.017]$ \\
5.66 -- 8.00 & 5.02 -- 6.32 & $0.019 [+0.013, -0.010]$ &
5.66 -- 8.00 & 6.32 -- 7.96 & $0.0051 [+0.0061, -0.0036]$ \\
5.66 -- 8.00 & 7.96 -- 10.02 & $0.0079 [+0.0050, -0.0037]$ &
5.66 -- 8.00 & 10.02 -- 12.62 & $0.021 [+0.0077, -0.0065]$ \\
5.66 -- 8.00 & 12.62 -- 15.89 & $0.030 [+0.010, -0.0089]$ &
5.66 -- 8.00 & 15.89 -- 20.00 & $0.032 [+0.053, -0.024]$ \\
8.00 -- 11.31 & 2.00 -- 2.52 & $2.4 [+1.9, -1.3]$ &
8.00 -- 11.31 & 2.52 -- 3.17 & $0.53 [+0.35, -0.27]$ \\
8.00 -- 11.31 & 3.17 -- 3.99 & $0.24 [+0.13, -0.10]$ &
8.00 -- 11.31 & 3.99 -- 5.02 & $0.050 [+0.038, -0.030]$ \\
8.00 -- 11.31 & 5.02 -- 6.32 & $0.027 [+0.018, -0.014]$ &
8.00 -- 11.31 & 6.32 -- 7.96 & $0.0048 [+0.0072, -0.0035]$ \\
8.00 -- 11.31 & 7.96 -- 10.02 & $0.019 [+0.0088, -0.0070]$ &
8.00 -- 11.31 & 10.02 -- 12.62 & $0.044 [+0.012, -0.010]$ \\
8.00 -- 11.31 & 12.62 -- 15.89 & $0.019 [+0.012, -0.0090]$ &
8.00 -- 11.31 & 15.89 -- 20.00 & $0.16 [+0.28, -0.12]$ \\
11.31 -- 16.00 & 2.00 -- 2.52 & $1.5 [+2.1, -1.1]$ &
11.31 -- 16.00 & 2.52 -- 3.17 & $1.7 [+0.68, -0.58]$ \\
11.31 -- 16.00 & 3.17 -- 3.99 & $0.21 [+0.19, -0.13]$ &
11.31 -- 16.00 & 3.99 -- 5.02 & $0.15 [+0.084, -0.068]$ \\
11.31 -- 16.00 & 5.02 -- 6.32 & $0.11 [+0.046, -0.040]$ &
11.31 -- 16.00 & 6.32 -- 7.96 & $0.011 [+0.015, -0.0081]$ \\
11.31 -- 16.00 & 7.96 -- 10.02 & $0.032 [+0.015, -0.013]$ &
11.31 -- 16.00 & 10.02 -- 12.62 & $0.034 [+0.016, -0.013]$ \\
11.31 -- 16.00 & 12.62 -- 15.89 & $0.018 [+0.018, -0.012]$ &
11.31 -- 16.00 & 15.89 -- 20.00 & $0.22 [+0.37, -0.16]$ \\
\hline
\end{tabular}
\end{table*}

\begin{table*}
\caption{Occurrence rates (number of planets per 100 stars) of our 10-by-10 bins of G-type stars (Fig.~\ref{fig:occ-rate-G}). Values are reported to two significant figures.}
\label{tab:TESS-SPOC 10x10 table G type}
\renewcommand{\arraystretch}{1.28}
\setlength{\tabcolsep}{4.1pt}
\begin{tabular}{ccc @{\ \vline\ } ccc}
\hline
Period ranges (days) & Radius ranges ($R_{\earth}$) & Occurrence rates (\%)
   & Period ranges (days) & Radius ranges ($R_{\earth}$) & Occurrence rates (\%) \\ 
\hline
0.50 -- 0.71 & 2.00 -- 2.52 & $0.0040 [+0.0063, -0.0030]$ &
0.50 -- 0.71 & 2.52 -- 3.17 & $0.0020 [+0.0025, -0.0014]$ \\
0.50 -- 0.71 & 3.17 -- 3.99 & $0.0014 [+0.0018, -0.0010]$ &
0.50 -- 0.71 & 3.99 -- 5.02 & $0.0024 [+0.0017, -0.0012]$ \\
0.50 -- 0.71 & 5.02 -- 6.32 & $0.00066 [+0.00099, -0.00049]$ &
0.50 -- 0.71 & 6.32 -- 7.96 & $0.00071 [+0.00088, -0.00049]$ \\
0.50 -- 0.71 & 7.96 -- 10.02 & $0.00096 [+0.00089, -0.00057]$ &
0.50 -- 0.71 & 10.02 -- 12.62 & $0.00044 [+0.00068, -0.00033]$ \\
0.50 -- 0.71 & 12.62 -- 15.89 & $0.00037 [+0.00061, -0.00028]$ &
0.50 -- 0.71 & 15.89 -- 20.00 & $0.0021 [+0.0034, -0.0015]$ \\
0.71 -- 1.00 & 2.00 -- 2.52 & $0.0048 [+0.0075, -0.0036]$ &
0.71 -- 1.00 & 2.52 -- 3.17 & $0.0018 [+0.0029, -0.0014]$ \\
0.71 -- 1.00 & 3.17 -- 3.99 & $0.0029 [+0.0029, -0.0019]$ &
0.71 -- 1.00 & 3.99 -- 5.02 & $0.0021 [+0.0019, -0.0013]$ \\
0.71 -- 1.00 & 5.02 -- 6.32 & $0.00058 [+0.00090, -0.00043]$ &
0.71 -- 1.00 & 6.32 -- 7.96 & $0.00037 [+0.00061, -0.00027]$ \\
0.71 -- 1.00 & 7.96 -- 10.02 & $0.00035 [+0.00058, -0.00026]$ &
0.71 -- 1.00 & 10.02 -- 12.62 & $0.00051 [+0.00070, -0.00037]$ \\
0.71 -- 1.00 & 12.62 -- 15.89 & $0.00092 [+0.0011, -0.00066]$ &
0.71 -- 1.00 & 15.89 -- 20.00 & $0.0078 [+0.0064, -0.0046]$ \\
1.00 -- 1.41 & 2.00 -- 2.52 & $0.0086 [+0.014, -0.0065]$ &
1.00 -- 1.41 & 2.52 -- 3.17 & $0.0094 [+0.0094, -0.0061]$ \\
1.00 -- 1.41 & 3.17 -- 3.99 & $0.0044 [+0.0047, -0.0029]$ &
1.00 -- 1.41 & 3.99 -- 5.02 & $0.0011 [+0.0017, -0.00081]$ \\
1.00 -- 1.41 & 5.02 -- 6.32 & $0.0012 [+0.0015, -0.00085]$ &
1.00 -- 1.41 & 6.32 -- 7.96 & $0.00069 [+0.00098, -0.00050]$ \\
1.00 -- 1.41 & 7.96 -- 10.02 & $0.00040 [+0.00063, -0.00029]$ &
1.00 -- 1.41 & 10.02 -- 12.62 & $0.00041 [+0.00066, -0.00031]$ \\
1.00 -- 1.41 & 12.62 -- 15.89 & $0.0040 [+0.0018, -0.0014]$ &
1.00 -- 1.41 & 15.89 -- 20.00 & $0.0042 [+0.0060, -0.0031]$ \\
1.41 -- 2.00 & 2.00 -- 2.52 & $0.034 [+0.030, -0.019]$ &
1.41 -- 2.00 & 2.52 -- 3.17 & $0.0092 [+0.012, -0.0065]$ \\
1.41 -- 2.00 & 3.17 -- 3.99 & $0.0029 [+0.0042, -0.0022]$ &
1.41 -- 2.00 & 3.99 -- 5.02 & $0.0035 [+0.0035, -0.0023]$ \\
1.41 -- 2.00 & 5.02 -- 6.32 & $0.0021 [+0.0022, -0.0014]$ &
1.41 -- 2.00 & 6.32 -- 7.96 & $0.0010 [+0.0014, -0.00074]$ \\
1.41 -- 2.00 & 7.96 -- 10.02 & $0.00059 [+0.00090, -0.00044]$ &
1.41 -- 2.00 & 10.02 -- 12.62 & $0.00094 [+0.0014, -0.00069]$ \\
1.41 -- 2.00 & 12.62 -- 15.89 & $0.0056 [+0.0023, -0.0018]$ &
1.41 -- 2.00 & 15.89 -- 20.00 & $0.0022 [+0.0036, -0.0016]$ \\
2.00 -- 2.83 & 2.00 -- 2.52 & $0.043 [+0.045, -0.027]$ &
2.00 -- 2.83 & 2.52 -- 3.17 & $0.048 [+0.024, -0.020]$ \\
2.00 -- 2.83 & 3.17 -- 3.99 & $0.014 [+0.011, -0.0085]$ &
2.00 -- 2.83 & 3.99 -- 5.02 & $0.0066 [+0.0059, -0.0041]$ \\
2.00 -- 2.83 & 5.02 -- 6.32 & $0.0068 [+0.0043, -0.0033]$ &
2.00 -- 2.83 & 6.32 -- 7.96 & $0.0041 [+0.0033, -0.0024]$ \\
2.00 -- 2.83 & 7.96 -- 10.02 & $0.0027 [+0.0023, -0.0016]$ &
2.00 -- 2.83 & 10.02 -- 12.62 & $0.0048 [+0.0035, -0.0026]$ \\
2.00 -- 2.83 & 12.62 -- 15.89 & $0.020 [+0.0050, -0.0044]$ &
2.00 -- 2.83 & 15.89 -- 20.00 & $0.0037 [+0.0062, -0.0028]$ \\
2.83 -- 4.00 & 2.00 -- 2.52 & $0.20 [+0.11, -0.089]$ &
2.83 -- 4.00 & 2.52 -- 3.17 & $0.071 [+0.047, -0.038]$ \\
2.83 -- 4.00 & 3.17 -- 3.99 & $0.035 [+0.021, -0.017]$ &
2.83 -- 4.00 & 3.99 -- 5.02 & $0.013 [+0.010, -0.0075]$ \\
2.83 -- 4.00 & 5.02 -- 6.32 & $0.014 [+0.0073, -0.0062]$ &
2.83 -- 4.00 & 6.32 -- 7.96 & $0.0084 [+0.0049, -0.0039]$ \\
2.83 -- 4.00 & 7.96 -- 10.02 & $0.0085 [+0.0046, -0.0037]$ &
2.83 -- 4.00 & 10.02 -- 12.62 & $0.031 [+0.0081, -0.0073]$ \\
2.83 -- 4.00 & 12.62 -- 15.89 & $0.072 [+0.011, -0.0096]$ &
2.83 -- 4.00 & 15.89 -- 20.00 & $0.0093 [+0.015, -0.0069]$ \\
4.00 -- 5.66 & 2.00 -- 2.52 & $0.59 [+0.26, -0.22]$ &
4.00 -- 5.66 & 2.52 -- 3.17 & $0.35 [+0.11, -0.093]$ \\
4.00 -- 5.66 & 3.17 -- 3.99 & $0.025 [+0.026, -0.017]$ &
4.00 -- 5.66 & 3.99 -- 5.02 & $0.038 [+0.018, -0.016]$ \\
4.00 -- 5.66 & 5.02 -- 6.32 & $0.018 [+0.011, -0.0087]$ &
4.00 -- 5.66 & 6.32 -- 7.96 & $0.015 [+0.0071, -0.0061]$ \\
4.00 -- 5.66 & 7.96 -- 10.02 & $0.0086 [+0.0058, -0.0046]$ &
4.00 -- 5.66 & 10.02 -- 12.62 & $0.047 [+0.0099, -0.0092]$ \\
4.00 -- 5.66 & 12.62 -- 15.89 & $0.038 [+0.011, -0.0095]$ &
4.00 -- 5.66 & 15.89 -- 20.00 & $0.032 [+0.055, -0.024]$ \\
5.66 -- 8.00 & 2.00 -- 2.52 & $1.1 [+0.46, -0.39]$ &
5.66 -- 8.00 & 2.52 -- 3.17 & $0.34 [+0.16, -0.14]$ \\
5.66 -- 8.00 & 3.17 -- 3.99 & $0.12 [+0.060, -0.049]$ &
5.66 -- 8.00 & 3.99 -- 5.02 & $0.031 [+0.022, -0.017]$ \\
5.66 -- 8.00 & 5.02 -- 6.32 & $0.021 [+0.015, -0.011]$ &
5.66 -- 8.00 & 6.32 -- 7.96 & $0.032 [+0.012, -0.010]$ \\
5.66 -- 8.00 & 7.96 -- 10.02 & $0.023 [+0.011, -0.0096]$ &
5.66 -- 8.00 & 10.02 -- 12.62 & $0.061 [+0.014, -0.013]$ \\
5.66 -- 8.00 & 12.62 -- 15.89 & $0.020 [+0.012, -0.0091]$ &
5.66 -- 8.00 & 15.89 -- 20.00 & $0.071 [+0.11, -0.054]$ \\
8.00 -- 11.31 & 2.00 -- 2.52 & $1.8 [+0.77, -0.65]$ &
8.00 -- 11.31 & 2.52 -- 3.17 & $0.96 [+0.29, -0.27]$ \\
8.00 -- 11.31 & 3.17 -- 3.99 & $0.21 [+0.12, -0.092]$ &
8.00 -- 11.31 & 3.99 -- 5.02 & $0.15 [+0.048, -0.044]$ \\
8.00 -- 11.31 & 5.02 -- 6.32 & $0.019 [+0.016, -0.011]$ &
8.00 -- 11.31 & 6.32 -- 7.96 & $0.014 [+0.011, -0.0079]$ \\
8.00 -- 11.31 & 7.96 -- 10.02 & $0.035 [+0.013, -0.011]$ &
8.00 -- 11.31 & 10.02 -- 12.62 & $0.068 [+0.016, -0.014]$ \\
8.00 -- 11.31 & 12.62 -- 15.89 & $0.0054 [+0.0088, -0.0040]$ &
8.00 -- 11.31 & 15.89 -- 20.00 & $0.34 [+0.54, -0.26]$ \\
11.31 -- 16.00 & 2.00 -- 2.52 & $1.6 [+0.86, -0.69]$ &
11.31 -- 16.00 & 2.52 -- 3.17 & $0.55 [+0.35, -0.27]$ \\
11.31 -- 16.00 & 3.17 -- 3.99 & $0.41 [+0.16, -0.14]$ &
11.31 -- 16.00 & 3.99 -- 5.02 & $0.029 [+0.041, -0.021]$ \\
11.31 -- 16.00 & 5.02 -- 6.32 & $0.12 [+0.040, -0.034]$ &
11.31 -- 16.00 & 6.32 -- 7.96 & $0.012 [+0.016, -0.0090]$ \\
11.31 -- 16.00 & 7.96 -- 10.02 & $0.020 [+0.016, -0.011]$ &
11.31 -- 16.00 & 10.02 -- 12.62 & $0.096 [+0.026, -0.021]$ \\
11.31 -- 16.00 & 12.62 -- 15.89 & $0.025 [+0.029, -0.017]$ &
11.31 -- 16.00 & 15.89 -- 20.00 & $0.45 [+0.75, -0.34]$ \\
\hline
\end{tabular}
\end{table*}

\begin{table*}
\caption{Occurrence rates (number of planets per 100 stars) of our 10-by-10 bins of K-type stars (Fig.~\ref{fig:occ-rate-K}). Values are reported to two significant figures.}
\label{tab:TESS-SPOC 10x10 table K type}
\renewcommand{\arraystretch}{1.28}
\setlength{\tabcolsep}{4.1pt}
\begin{tabular}{ccc @{\ \vline\ } ccc}
\hline
Period ranges (days) & Radius ranges ($R_{\earth}$) & Occurrence rates (\%)
   & Period ranges (days) & Radius ranges ($R_{\earth}$) & Occurrence rates (\%) \\ 
\hline
0.50 -- 0.71 & 2.00 -- 2.52 & $0.0038 [+0.0062, -0.0028]$ &
0.50 -- 0.71 & 2.52 -- 3.17 & $0.0021 [+0.0035, -0.0016]$ \\
0.50 -- 0.71 & 3.17 -- 3.99 & $0.0018 [+0.0030, -0.0013]$ &
0.50 -- 0.71 & 3.99 -- 5.02 & $0.0014 [+0.0023, -0.0010]$ \\
0.50 -- 0.71 & 5.02 -- 6.32 & $0.0016 [+0.0023, -0.0012]$ &
0.50 -- 0.71 & 6.32 -- 7.96 & $0.0023 [+0.0028, -0.0016]$ \\
0.50 -- 0.71 & 7.96 -- 10.02 & $0.0013 [+0.0020, -0.00098]$ &
0.50 -- 0.71 & 10.02 -- 12.62 & $0.0018 [+0.0027, -0.0013]$ \\
0.50 -- 0.71 & 12.62 -- 15.89 & $0.0020 [+0.0030, -0.0015]$ &
0.50 -- 0.71 & 15.89 -- 20.00 & $0.011 [+0.019, -0.0086]$ \\
0.71 -- 1.00 & 2.00 -- 2.52 & $0.017 [+0.016, -0.010]$ &
0.71 -- 1.00 & 2.52 -- 3.17 & $0.0058 [+0.0087, -0.0043]$ \\
0.71 -- 1.00 & 3.17 -- 3.99 & $0.0034 [+0.0046, -0.0024]$ &
0.71 -- 1.00 & 3.99 -- 5.02 & $0.0084 [+0.0054, -0.0039]$ \\
0.71 -- 1.00 & 5.02 -- 6.32 & $0.0018 [+0.0029, -0.0013]$ &
0.71 -- 1.00 & 6.32 -- 7.96 & $0.0018 [+0.0030, -0.0013]$ \\
0.71 -- 1.00 & 7.96 -- 10.02 & $0.0015 [+0.0026, -0.0011]$ &
0.71 -- 1.00 & 10.02 -- 12.62 & $0.0067 [+0.0047, -0.0036]$ \\
0.71 -- 1.00 & 12.62 -- 15.89 & $0.0031 [+0.0047, -0.0023]$ &
0.71 -- 1.00 & 15.89 -- 20.00 & $0.024 [+0.039, -0.018]$ \\
1.00 -- 1.41 & 2.00 -- 2.52 & $0.018 [+0.021, -0.012]$ &
1.00 -- 1.41 & 2.52 -- 3.17 & $0.014 [+0.015, -0.0093]$ \\
1.00 -- 1.41 & 3.17 -- 3.99 & $0.019 [+0.012, -0.0094]$ &
1.00 -- 1.41 & 3.99 -- 5.02 & $0.0071 [+0.0080, -0.0048]$ \\
1.00 -- 1.41 & 5.02 -- 6.32 & $0.0021 [+0.0035, -0.0016]$ &
1.00 -- 1.41 & 6.32 -- 7.96 & $0.0019 [+0.0031, -0.0014]$ \\
1.00 -- 1.41 & 7.96 -- 10.02 & $0.0018 [+0.0030, -0.0014]$ &
1.00 -- 1.41 & 10.02 -- 12.62 & $0.0020 [+0.0034, -0.0015]$ \\
1.00 -- 1.41 & 12.62 -- 15.89 & $0.0078 [+0.0056, -0.0038]$ &
1.00 -- 1.41 & 15.89 -- 20.00 & $0.018 [+0.030, -0.014]$ \\
1.41 -- 2.00 & 2.00 -- 2.52 & $0.031 [+0.039, -0.022]$ &
1.41 -- 2.00 & 2.52 -- 3.17 & $0.019 [+0.021, -0.013]$ \\
1.41 -- 2.00 & 3.17 -- 3.99 & $0.010 [+0.012, -0.0067]$ &
1.41 -- 2.00 & 3.99 -- 5.02 & $0.0046 [+0.0071, -0.0033]$ \\
1.41 -- 2.00 & 5.02 -- 6.32 & $0.0073 [+0.0064, -0.0043]$ &
1.41 -- 2.00 & 6.32 -- 7.96 & $0.0028 [+0.0044, -0.0021]$ \\
1.41 -- 2.00 & 7.96 -- 10.02 & $0.0033 [+0.0052, -0.0024]$ &
1.41 -- 2.00 & 10.02 -- 12.62 & $0.0076 [+0.0063, -0.0040]$ \\
1.41 -- 2.00 & 12.62 -- 15.89 & $0.0077 [+0.0074, -0.0046]$ &
1.41 -- 2.00 & 15.89 -- 20.00 & $0.013 [+0.021, -0.0094]$ \\
2.00 -- 2.83 & 2.00 -- 2.52 & $0.10 [+0.11, -0.068]$ &
2.00 -- 2.83 & 2.52 -- 3.17 & $0.12 [+0.057, -0.048]$ \\
2.00 -- 2.83 & 3.17 -- 3.99 & $0.015 [+0.019, -0.011]$ &
2.00 -- 2.83 & 3.99 -- 5.02 & $0.017 [+0.013, -0.0093]$ \\
2.00 -- 2.83 & 5.02 -- 6.32 & $0.010 [+0.0090, -0.0059]$ &
2.00 -- 2.83 & 6.32 -- 7.96 & $0.0076 [+0.0076, -0.0046]$ \\
2.00 -- 2.83 & 7.96 -- 10.02 & $0.0036 [+0.0058, -0.0026]$ &
2.00 -- 2.83 & 10.02 -- 12.62 & $0.048 [+0.016, -0.014]$ \\
2.00 -- 2.83 & 12.62 -- 15.89 & $0.021 [+0.014, -0.011]$ &
2.00 -- 2.83 & 15.89 -- 20.00 & $0.016 [+0.026, -0.012]$ \\
2.83 -- 4.00 & 2.00 -- 2.52 & $0.18 [+0.14, -0.11]$ &
2.83 -- 4.00 & 2.52 -- 3.17 & $0.13 [+0.075, -0.058]$ \\
2.83 -- 4.00 & 3.17 -- 3.99 & $0.017 [+0.025, -0.012]$ &
2.83 -- 4.00 & 3.99 -- 5.02 & $0.031 [+0.025, -0.018]$ \\
2.83 -- 4.00 & 5.02 -- 6.32 & $0.079 [+0.027, -0.023]$ &
2.83 -- 4.00 & 6.32 -- 7.96 & $0.018 [+0.013, -0.0092]$ \\
2.83 -- 4.00 & 7.96 -- 10.02 & $0.013 [+0.013, -0.0083]$ &
2.83 -- 4.00 & 10.02 -- 12.62 & $0.10 [+0.028, -0.024]$ \\
2.83 -- 4.00 & 12.62 -- 15.89 & $0.025 [+0.020, -0.015]$ &
2.83 -- 4.00 & 15.89 -- 20.00 & $0.045 [+0.070, -0.034]$ \\
4.00 -- 5.66 & 2.00 -- 2.52 & $0.72 [+0.36, -0.29]$ &
4.00 -- 5.66 & 2.52 -- 3.17 & $0.52 [+0.19, -0.16]$ \\
4.00 -- 5.66 & 3.17 -- 3.99 & $0.056 [+0.052, -0.035]$ &
4.00 -- 5.66 & 3.99 -- 5.02 & $0.023 [+0.025, -0.016]$ \\
4.00 -- 5.66 & 5.02 -- 6.32 & $0.026 [+0.024, -0.017]$ &
4.00 -- 5.66 & 6.32 -- 7.96 & $0.060 [+0.029, -0.023]$ \\
4.00 -- 5.66 & 7.96 -- 10.02 & $0.049 [+0.026, -0.021]$ &
4.00 -- 5.66 & 10.02 -- 12.62 & $0.037 [+0.022, -0.017]$ \\
4.00 -- 5.66 & 12.62 -- 15.89 & $0.059 [+0.029, -0.022]$ &
4.00 -- 5.66 & 15.89 -- 20.00 & $0.12 [+0.18, -0.087]$ \\
5.66 -- 8.00 & 2.00 -- 2.52 & $1.1 [+0.48, -0.41]$ &
5.66 -- 8.00 & 2.52 -- 3.17 & $0.71 [+0.28, -0.25]$ \\
5.66 -- 8.00 & 3.17 -- 3.99 & $0.12 [+0.091, -0.072]$ &
5.66 -- 8.00 & 3.99 -- 5.02 & $0.11 [+0.055, -0.044]$ \\
5.66 -- 8.00 & 5.02 -- 6.32 & $0.028 [+0.025, -0.016]$ &
5.66 -- 8.00 & 6.32 -- 7.96 & $0.014 [+0.018, -0.0098]$ \\
5.66 -- 8.00 & 7.96 -- 10.02 & $0.10 [+0.039, -0.030]$ &
5.66 -- 8.00 & 10.02 -- 12.62 & $0.056 [+0.030, -0.023]$ \\
5.66 -- 8.00 & 12.62 -- 15.89 & $0.013 [+0.022, -0.0096]$ &
5.66 -- 8.00 & 15.89 -- 20.00 & $0.063 [+0.10, -0.047]$ \\
8.00 -- 11.31 & 2.00 -- 2.52 & $2.0 [+0.78, -0.65]$ &
8.00 -- 11.31 & 2.52 -- 3.17 & $1.3 [+0.44, -0.38]$ \\
8.00 -- 11.31 & 3.17 -- 3.99 & $0.18 [+0.12, -0.094]$ &
8.00 -- 11.31 & 3.99 -- 5.02 & $0.060 [+0.058, -0.039]$ \\
8.00 -- 11.31 & 5.02 -- 6.32 & $0.020 [+0.027, -0.014]$ &
8.00 -- 11.31 & 6.32 -- 7.96 & $0.015 [+0.021, -0.011]$ \\
8.00 -- 11.31 & 7.96 -- 10.02 & $0.054 [+0.031, -0.023]$ &
8.00 -- 11.31 & 10.02 -- 12.62 & $0.048 [+0.031, -0.023]$ \\
8.00 -- 11.31 & 12.62 -- 15.89 & $0.044 [+0.070, -0.032]$ &
8.00 -- 11.31 & 15.89 -- 20.00 & $0.49 [+0.79, -0.36]$ \\
11.31 -- 16.00 & 2.00 -- 2.52 & $1.4 [+1.1, -0.83]$ &
11.31 -- 16.00 & 2.52 -- 3.17 & $1.6 [+0.64, -0.56]$ \\
11.31 -- 16.00 & 3.17 -- 3.99 & $0.28 [+0.20, -0.15]$ &
11.31 -- 16.00 & 3.99 -- 5.02 & $0.076 [+0.081, -0.051]$ \\
11.31 -- 16.00 & 5.02 -- 6.32 & $0.070 [+0.069, -0.045]$ &
11.31 -- 16.00 & 6.32 -- 7.96 & $0.059 [+0.052, -0.036]$ \\
11.31 -- 16.00 & 7.96 -- 10.02 & $0.090 [+0.070, -0.052]$ &
11.31 -- 16.00 & 10.02 -- 12.62 & $0.12 [+0.071, -0.055]$ \\
11.31 -- 16.00 & 12.62 -- 15.89 & $0.037 [+0.059, -0.027]$ &
11.31 -- 16.00 & 15.89 -- 20.00 & $0.71 [+1.2, -0.53]$ \\
\hline
\end{tabular}
\end{table*}


\bsp
\label{lastpage}
\end{document}